\RequirePackage[2018-12-01]{latexrelease}
\documentclass{gGAF2e}
\usepackage{graphicx}
\usepackage{epstopdf, epsfig}
\usepackage{bbm}
\usepackage{color}
\usepackage{comment}
\usepackage[normalem]{ulem}

\begin{document}

\jvol{00} \jnum{00} \jyear{2012} 

\markboth{Wareing et al.}{Data-driven properties of convection}

\articletype{ARTICLE}

\title{Data-driven discovery of the equations of turbulent convection}

\author{Christopher J. Wareing${\dag}$$^{\ast}$\thanks{$^\ast$Corresponding author. Email: C.J.Wareing@leeds.ac.uk
\vspace{6pt}}, 
Alasdair T. Roy${\dag}$, 
Matthew Golden${\ddag}$,
Roman O. Grigoriev${\ddag}$,
and  Steven M. Tobias${\dag}$\\ 
\vspace{6pt}  ${\dag}$ Department of Applied Mathematics, School of Mathematics, University of Leeds, Leeds, LS2 9JT, UK\\
\vspace{6pt}  ${\ddag}$ School of Physics, Georgia Institute of Technology, North Avenue, Atlanta, GA 30332, USA\\
\vspace{6pt}\received{recv. 2024 November 14; acc. 2025 May 14} }

\maketitle

\begin{abstract}
We compare the efficiency and ease-of-use of
the 
Sparse Identification of Nonlinear Dynamics (SINDy) algorithm
and Sparse Physics-Informed Discovery of Empirical Relations (SPIDER) framework
in recovering the relevant governing equations and boundary conditions
from data generated by direct numerical simulations (DNS) of turbulent convective flows.
In the former case, a weak-form implementation pySINDy is used. 
Time-dependent data for two- (2D) and three-dimensional (3D) DNS simulation of 
Rayleigh-B\'enard convection and convective plane Couette flow is generated using the Dedalus PDE framework for spectrally solving differential equations.  Using pySINDy
we are able to recover the governing equations of 2D models of Rayleigh-B\'enard convection at 
Rayleigh numbers, $R$, from laminar, through transitional to moderately turbulent flow conditions,
albeit with increasing difficulty with larger Rayleigh number, especially in 
recovery of the diffusive terms (with coefficient magnitude proportional to $\sqrt{1/R}$).
SPIDER requires a much smaller library of terms and 
we are able to  recover more easily the governing equations for a wider range of $R$ in 2D and 3D convection and plane flow models 
and go on to recover 
constraints (the incompressibility condition) and boundary conditions, demonstrating the 
benefits and capabilities 
of SPIDER to go beyond pySINDy for these fluid problems governed by second-order PDEs.
At the highest values of $R$, discrepancies appear between the governing equations that are solved and those that are discovered by SPIDER. We
find that this is likely associated with limited resolution of DNS, demonstrating the potential of machine-learning methods to
validate numerical solvers and solutions for such flow problems.
We also find that properties of the flow, specifically the correlation time and spatial
scales, should inform the initial selection of spatiotemporal subdomain sizes for both 
pySINDy and SPIDER. Adopting this default position has the potential to reduce trial and error in selection of data parameters, saving considerable
time and effort and allowing the end user of these or similar methods to focus on the
importance of setting the power of the integrating polynomial in these weak-form methods and the
tolerance of the optimiser technique selected.


\end{abstract}

\section{Introduction}

Geophysical and astrophysical flows are characterised by their nonlinear interaction over 
a vast range of spatial and temporal scales and hence the presence of turbulence. Often 
this turbulence is inhomogeneous and anisotropic owing to the nature of the driving 
(and possibly the presence of rotation, stratification and mean flows) \citep{Marston_tobias_2023}. 
The large range of scales often requires the modelling of scales that are not captured 
by the numerical solution, sometimes denoted as subgrid modelling. As recently reviewed
in \cite{moser21} this can take many forms,
including the derivation of statistical representations of the unresolved scales
(e.g. \citet{chorin02}), 
utilising machine learning to replicate the response of the unresolved scales 
\citep{bolton19,zanna20} or discovering the effective equations satisfied by the low-order 
statistics of those scales. These `turbulent closure equations' or parameterisations of 
the turbulent transport coefficients are the subject of much ongoing cutting-edge research; 
see the summary in \cite{jak24}.

In this paper,
as a first step to test the methods that we intend in future work to apply to
the search for turbulent closure models \citep[such as][]{gar10},
we focus on the simpler problem of the reconstruction of the relevant full underlying
equations from data. This problem is easier because it does not rely on assuming the correct
definition of an averaging procedure or maintaining realisability of the reconstructed equations. 
Here we are interested in how two leading equation inference methods perform for a turbulent flow driven by 
buoyancy (and, in some cases, driven from the boundaries). This form of driving is important in 
many geophysical and astrophysical settings where thermal convection is responsible for heat 
(and possibly momentum) transport in planets and stars.

Whilst recovery of evolution equations via some form of regression is now common, with
the Sparse Identification of Nonlinear Dynamics (SINDy) algorithm \citep{brun16} so far 
proven to be the default choice of approach, identification of flow constraints and boundary 
conditions from data - simulated or experimental - is still relatively rare. Reduction
techniques and regressive reconstruction, using neural networks, have been shown to be able
to reconstruct the latent fluid state for a number of unsteady flows, e.g. vortex shedding
at Reynolds numbers up to 20000 \citep{dubois22}. Similarly, artificial neural networks
have been trained to solve a non-linear regression problem and obtain inlet conditions
for a fluid flow problem; see \cite{veras21} and citations of this paper using this strategy.
As far as we are aware, sparse regression machine-learning techniques have only recently 
been shown as able to recover equations of flow constraints and boundary conditions through
the Sparse Physics-Informed Discovery of Empirical Relations (SPIDER) framework
\citep{gure21}.

In this paper, as a first step toward turbulence closure model recovery,
we apply both SINDy and SPIDER to convective flow problems. The goal is to
compare their performance and relative merits and look to provide guidance for the end-user 
of either technique. This paper is organised as follows:
In Section \ref{models}, we present the convection models we have considered.
In Section \ref{nummeth} we obtain simulated data of these models using spectral methods.
We introduce and compare the two equation discovery methods, noting their similar objectives
and highlighting the differences in application in Section \ref{mlmeth}. We explore the
equation recovery performance of both methods in Section \ref{results} and discuss
the limitations, possible reasons for the differences and some advice for future users
of either machine-learning approach in Section \ref{discuss}. Finally, we summarise
this work and look to future directions in Section \ref{conc}.

\section{Models}\label{models}

\subsection{Rayleigh-B\'enard convection}

We consider classical Rayleigh-B\'enard convective fluid flow in a 
flat horizontally-periodic
layer heated from below. We assume the Boussinesq approximation.
The non-dimensionalized dynamical equations for the fluid are
\begin{subequations}\label{RBeqns}
\begin{equation}
\frac{\upartial{\bm{u}}}{\upartial t}
\,+\,({\bm{u}}{\bm{\cdot}}
{\bm{\nabla}})\,{\bm{u}}\,=
\,-\,{\bm{\nabla}}p
\,+\,T\,\bm{\hat{z}}
\,+\,\sqrt{\frac{P}{R}}\,\nabla^2{\bm{u}}\,,
\label{RB-NSeqn}
\end{equation}
\begin{equation}
\frac{\upartial{T}}{\upartial t}
\,+\,({\bm{u}}{\bm{\cdot}}
{\bm{\nabla}})\,{T}\,=
\,\frac{1}{\sqrt{R\,P}}\,\nabla^2{T}\,,
\label{RB-Teqn}
\end{equation}
\begin{equation}
{\bm{\nabla}}\,{\bm{\cdot}}\,{\bm{u}}\,=
\,0\,,
\label{RB-cont}
\end{equation}
\end{subequations}
where  the non-dimensional fields are $\bm{u}$, the vector velocity field, the scalar pressure field $p$  and  the 
scalar temperature field $T$. Here the non-dimensional parameters are the Rayleigh number $R=\alpha g(\Delta T)L^{3}/\nu\,\kappa$ and the Prandtl number $P=\nu/\kappa$.  Gravity acts in the  $-\hat{z}$ direction and $\alpha$, $\kappa$ 
and $\nu$ are the thermal heat expansion coefficient, thermal diffusivity and kinematic viscosity 
respectively, $g$ is the acceleration due to gravity, $L$ is the vertical
height of the layer, and $\Delta T$ is the temperature difference 
between the bottom and top plates. 

For the non-dimensionalization we have used $L$ as the 
length-scale, the buoyancy free-fall velocity ($\sqrt{\alpha g (\Delta T) L}$) as the velocity scale,
and $\Delta T$ as the temperature scale. As a consequence, $L/\sqrt{\alpha g (\Delta T) L}$
is the time-scale. We consider both two-dimensional and three-dimensional cases.
We use an aspect ratio of four for the layer, such that for the
2D simulations the layer is four times longer in the $x$ direction than in the
buoyancy direction of $z$, and for the 3D simulations the layer is four times
longer in both the $x$ and $y$ directions than in the 
vertical $z$ direction, defined as the one opposite to gravity. We confine
ourselves in this study to Prandtl number $P\,=\,1$. 

We impose the stress-free, isothermal, boundary condition on the problem, such that fluid conditions on
the top ($z=1$) and bottom ($z=0$) boundaries satisfy the following relations as necessary
depending on the number of dimensions
\begin{subequations}
    \begin{align}
        \frac{\upartial{u_x}}{\upartial z}\,&=\,0\,,\\
        \frac{\upartial{u_y}}{\upartial z}\,&=\,0\,,\\
        u_z\,&=\,0\,,\\
        T(z=0)\,&=\,1\,,\\
        T(z=1)\,&=\,0\,.
    \end{align}
\end{subequations}

\subsection{Planar convective Couette flow}

In this 2D model of planar convective Couette flow, we extend the model of Rayleigh-B\'enard 
convection above to move the top and bottom boundaries in antiparallel directions, such that
the non-dimensionalized dynamical equations and the boundary condition on the temperature 
remain unchanged, but the boundary condition on the velocity is now that of the no-slip condition
\begin{subequations}
    \begin{align}
        u_z(z=0)\,&=\,0\,,\\
        u_z(z=1)\,&=\,0\,,\\
        u_x(z=0)\,&=\,-\rm{U}_0\,,\\
        u_x(z=1)\,&=\,\rm{U}_0\,,
    \end{align}
\end{subequations}
while the boundary condition on the temperature remains unchanged. $\rm{U}_0$ therefore represents the velocity on the the walls measured in units of the free-fall velocity. The vertical 
height and aspect ratio of the problem remain the same, resulting in a physical 
extent of $4 \times 1$ in this 2D problem.

\section{Numerical method and Resolution}\label{nummeth}

The above models are solved numerically for a Cartesian geometry using the Dedalus framework for 
solving partial differential equations using spectral methods, specifically version
3.0.1\footnote{https://github.com/DedalusProject/dedalus/releases/tag/v3.0.1} \citep{dedalus3}. The 2D model 
of Rayleigh-B\'enard convection closely follows that of the 2D initial value problem (IVP) in the Cartesian 
examples packaged in the examples subdirectory of the Dedalus3 code repository, except that
the stress-free, rather than no-slip, boundary condition is imposed on the problem. Furthermore, rather than the second-order, two-stage DIRK+ERK (d3.RK222) timestepping scheme
(\citealt{asch97} sec. 2.6), 
we use the third-order, four-stage DIRK+ERK (d3.RK443) scheme (\citealt{asch97} sec. 2.8) 
for greater accuracy and stability.

We use a periodic Fourier basis for the $x$ and $y$ directions,
and a Chebyshev basis for the $z$ direction. In all bases, a 
Dedalus dealiasing factor of $3/2$ (equivalent to the conventional $2/3$) is
employed. Scalar fields are defined for pressure and temperature and a vector field for velocity.
The {\it generalised tau method} is used for imposing boundary conditions; explicit tau terms are added to the PDE introducing degrees of freedom
that allow the problem to be solved exactly over polynomials. 
Further details can be obtained
from the Dedalus repository$^{1}$ and in the Dedalus methods paper \citep{dedalus3}.

Our simulations use a Courant-Friedrichs-Lewy solver (d3.CFL) with an initial and maximum timestep
set to 0.01, a safety factor (the Courant number) of 0.5, an iteration cadence of 10 for calculating 
a new timestep, a maximum and minimum fractional change between timesteps of 1.5 and 0.5 respectively
and a fractional change threshold for changing the timestep of 0.05. We have not explored variation
of these parameters.

We monitor Reynolds number, Nusselt number, average kinetic energy and total kinetic 
energy in order to decide when a statistically steady state has been reached. 
We arbitrarily select this regime for the application of machine learning methods,
but one could examine the ability in other more transient periods of the flow.

It is common knowledge that, in order for a data-driven approach to recover a mathematical 
model, the data must exhibit enough variation on both space and time to sample the state of 
the physical problem \citep{scha18}. For this reason, we employ a constant high resolution
for each instance of the Rayleigh number, as detailed in this section. The use of the same resolution for all, defined so as to resolve the turbulent evolution and structure 
at the largest $R$, may seem over-resolved and numerically
expensive, but tests show there are no detrimental effects compared to lower resolutions and 
all run to a statistically steady state in a matter of hours at most on single 40-core blades of the HPC resources 
for the 2D problem and ten 40-core blades for the 3D problem.
We include our Dedalus scripts for all the models studied in
the data repository accompanying this paper for reproducibility and
replicability\footnote{https://doi.org/10.5518/1577}.

\subsection{2D Rayleigh-B\'enard convection}

In two dimensions, we perform simulations for Rayleigh
numbers $R$\,=\,10$^6$, 10$^8$, 10$^{10}$ and 10$^{12}$, modelling the laminar, transitional
and turbulent regimes, as confirmed by monitoring of the Nusselt and Reynolds numbers.
The physical domain is discretized into a uniform grid of $1024$ points in the $x$ direction 
and $384$ non-uniform points in the $z$ direction defined by the Chebyshev basis. 
The Chebyshev basis allows
for more accurate resolution of the boundary layers than a uniformly spaced  grid
of 384 points. This vertical resolution places a similar number of grid
points in the boundary layer compared to other DNS studies of the same
problem \citep{zhu18}. Power spectra of the whole flow also demonstrate
that the steep drop off in power at small scales is fully captured  
at this resolution for $R$ values of $10^6$ and $10^8$, but then approaching
the limit at $10^{10}$ and at the limit for $10^{12}$. We return to the 
importance of resolution and how this
may be affecting the methods in later discussions.

The initial physical values of the scalar pressure field and components of the vector velocity 
field are all set to zero. The physical temperature (buoyancy) field is filled with random noise 
on a normal distribution curve with a magnitude of $10^{-5}$. The noise is damped
at the walls according to $T(z) \to T(z) \times z \times (L-z)$ and a linear background is imposed
$T(z) = T(z) + (L-z)$.

\begin{figure}
\centering
\includegraphics[width=1\linewidth]{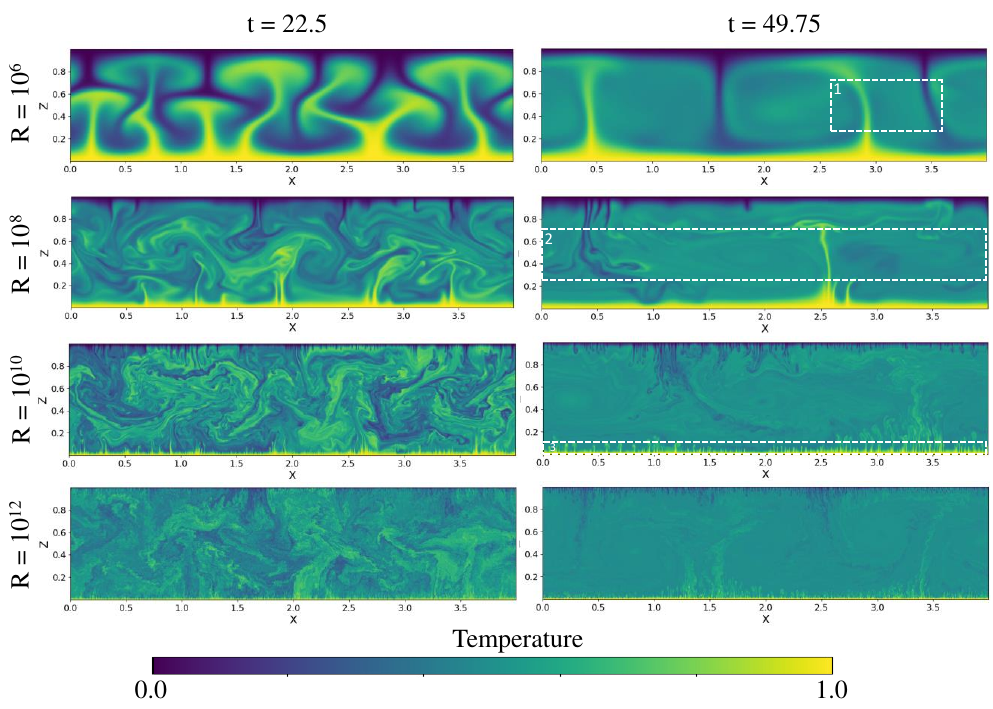}
\caption{Snapshots of 2D Rayleigh-B\'enard convection at varying times and Rayleigh number. 
Snapshots are shown early in the initial evolution ($t=22.5$ in the left column) and once
the steady state has become established ($t=49.75$ in the right column).
Boxes indicated by numbers 1, 2 and 3 overlaid in the right column show the approximate
sample domains for the various pySINDy tests discussed in Section \ref{results}.}
\label{figure1}
\end{figure}

The simulations are evolved until a time $t=50$, which our investigations have shown is the time
at which all simulations have reached an approximately steadily evolving saturated state. Note that some
simulations, with lower Rayleigh numbers, reach a steadily evolving state of convective rolls sooner, as is to be expected.
At $t=50$ the Dedalus model is restarted 
and evolved until $t=60$ with
the state of the fluid checkpointed  at a much higher cadence, $\Delta t = 0.01$, 
in order to provide the resolution in time required for the machine learning approaches.
We show snapshots of the temperature  for a range of parameters in Figure \ref{figure1}.

\subsection{3D Rayleigh B\'enard convection }

In three dimensions, we perform simulations 
for Rayleigh numbers $R$\,=\,10$^4$, 10$^5$, 10$^6$ and 10$^7$, again modelling the same range of
fluid regimes and monitored by Nusselt and Reynolds numbers. The physical domain is discretized 
into uniform grids of $128$ points in the $x$ and $y$ directions and $192$ non-uniform points 
in the $z$ direction defined by the Chebyshev basis.

The initial values of the fields are generated in the same manner as in the 2D model but here
the simulations are evolved considerably further to a time of $t=150$ in order to reach a statistically steady
 state, before restarting and evolving the simulation up to $t=175$ in order to obtain 
snapshots of the fluid state at the smaller interval of $\Delta t = 0.01$ to provide high cadence 
snapshots of the fluid for the machine learning algorithms.

\subsection{Two-dimensional planar convective Couette flow}

We perform simulations for 
the Rayleigh number $R$\,=\,10$^8$, principally to examine the capability of machine
learning methods for recovering a boundary condition involving a moving boundary. The physical domain is discretized 
into a uniform grid of $1024$ points in the $x$ direction and $384$ non-uniform points 
in the $z$ direction defined by the Chebyshev basis. The initial values of the fields are
again generated in the same manner as in the 2D Rayleigh-B\'enard convection model, with 
vertical boundary velocities of U$_0$ = 0.5 and 1.0. The simulations are evolved up
to $t=150$ and then restarted in order to obtain fluid data at the same smaller interval of
$\Delta t = 0.01$, up to $t=160$.

\section{Equation discovery methods}\label{mlmeth}

In this section we introduce the equation discovery algorithms and methods we have used,
along with precise details of how we have applied them to the data obtained above.
The methods we use are the pySINDy implementation of the Sparse Identification of 
Nonlinear Dynamics (SINDy) algorithm
and the Sparse Physics-Informed Discovery of Empirical Relations (SPIDER) framework.
For a recent review of machine learning for PDEs with pySINDy see \cite{brun23} and the 
references therein. The application of the SPIDER algorithm to turbulent channel flow
\citep{gure21} also provides a wide introduction to the history and application of
machine learning to PDEs. For reproducibility and completeness, we include our pySINDy and 
SPIDER analysis scripts, as well as the instances of the libraries used, in the 
repository accompanying this paper\footnote{https://doi.org/10.5518/1577}.

\subsection{The SINDy algorithm}

Model discovery is an active field of research which aims to discover models
of systems and can find first-principles equations that govern such systems 
by using observational data. SINDy \citep{brun16} has
become a popular approach to this problem as it allows the discovery of
sparse models which represent the data with a method that is both fast and simple to
implement. For this reason SINDy has seen many extensions, such as: implicit equation
identification \citep{kahe2020}, latent space dynamics \citep{cham2019, gao2024},
globally stable models \citep{kapt2021} and rapid-Bayesian formulations 
\citep{fung2024}. It has since been made available publicly in an open-source 
Python package called pySINDy \citep{silv2020} as a generalisable framework for 
solving sparse regression problems featuring many of these extensions. 

To formulate the optimisation problem in pySINDy, we aim to produce a linear
combination of candidate features from a library of terms which generates the dynamics. Explicitly, 
given $n$ measurements of a $d$ dimensional problem, we can construct a state 
matrix $\boldsymbol{X} \in \mathbb{R}^{n\times d}$ where the columns 
$j \in [0, 1, \dots, d]$ of $\boldsymbol{X}$ represent the state variables 
and the rows $i \in [0, 1, \dots, n]$ give measurements in time for each state. 
The regression problem is cast by solving 
\begin{equation}
    \dot{\boldsymbol{X}}_j= \boldsymbol{\Theta(\boldsymbol{X})} \boldsymbol{\xi}_j
\end{equation}
where $\boldsymbol{\Theta} \in \mathbb{R}^{n \times p}$ is a feature library
constructed of $p$ terms, $\boldsymbol{\xi}_j$ is the coefficient vector for 
the state variable $j$ and $\dot{\boldsymbol{X}}_j$ is the time-derivative of 
the state variable. The feature library can then be constructed from any terms 
the user wishes to include, such as polynomial terms of the input states. 

Following its introduction, the SINDy algorithm was later extended 
to learn PDEs from data in the PDE-FIND algorithm \citep{rudy17}. Given some 
spatiotemporal data, the extension from ODE to PDE identification is simple. For a general
PDE of the form
\begin{equation} \label{eqn:general_pde_form}
    \partial_t \boldsymbol{u} = \sum_{l=0}^N c_l \boldsymbol{f}_l (\boldsymbol{u}, \partial_t \boldsymbol{u}, \partial_t^2 \boldsymbol{u}, \nabla \boldsymbol{u}, \nabla^2 \boldsymbol{u},...) = 0
\end{equation}
where $c_l$ is the coefficient of $\boldsymbol{f}_l$, we construct a state matrix
$\boldsymbol{U}$ where the columns are vectors of length $m \times n$ where $n$ 
is the number of temporal measurements and $n$ the number of spatial measurements. 
In our feature library we now include spatial derivative terms such as $\boldsymbol{U}_{xx}$ and $\boldsymbol{U} \boldsymbol{U}_x$ to represent 
typical PDE terms. The regression problem is then 
\begin{equation} \label{eqn:PDE_find}
    \boldsymbol{U}_{t,j} = \boldsymbol{\Theta}(\boldsymbol{U}) \boldsymbol{\xi}_j
\end{equation}
where subscript $t,j$ represents the time derivative of $\boldsymbol{U}$ for the $j^{th}$ state. 


A primary limitation of SINDy as expressed in Equation (\ref{eqn:PDE_find}) is 
the use of finite-differencing in approximating spatiotemporal derivatives. 
Even extremely low amplitudes of noise have been noted to cause substantial errors 
in the identification of PDEs \citep{rudy17}. This issue is further 
compounded by the presence of nonlinear product terms in the feature library 
which conspire to violate the linear regression assumption of additive Gaussian noise
\citep{fung2024}. The effect of noise can be ameliorated by using the weak form of PDEs \citep{gure19}. In particular, a weak form of SINDy was introduced by \citet{mess21}.

The advantage of the weak form lies in the ability to transfer high-order 
derivatives from the state variables $\boldsymbol{u}$ on to a weight function
$\boldsymbol{w}$ with known form and well defined derivatives. Equation
(\ref{eqn:general_pde_form}) is multiplied by $\boldsymbol{w}$ and each term 
in the library is integrated over $K$ randomly sampled subdomains of volume
$\Omega_k$ for $k\in [1,2, \dots, K]$. The optimisation is then expressed as
\begin{equation} \label{eqn:weak_sindy_regression}
    \boldsymbol{q}_{0,j} = \sum_{n=1}^N c_{n,j} \boldsymbol{q}_n = Q \boldsymbol{c}_j 
\end{equation}
where $Q = [\boldsymbol{q}_1, ... ,\boldsymbol{q}_N]$ is the collection of $N$
different features of the PDE with the form
\begin{equation}
    \boldsymbol{q}^k_n = \int_{\Omega_k} \boldsymbol{w} \cdot \boldsymbol{f}_n \text{d} \Omega_k.  
\end{equation}
We can then perform PDE identification by assuming the form of the PDE is not 
known, and substituting a selection of candidate function $\boldsymbol{\Theta}_m$ 
for $\boldsymbol{f}_n$. This casts Equation (\ref{eqn:weak_sindy_regression}) as a
linear regression problem; i.e. we wish to  minimise the residual sum of 
squared errors of the $K$ different integrals on each subdomain. 
The regression is performed over the results of all the $K$ integrated subdomains at once. 
Both the number $K$ of subdomains used and their locations impact the results of 
regression as has been shown previously \citep{abram22,gure21}. A larger number of 
subdomains ensures that the data are more diverse and representative, yielding more 
structurally robust results for the functional form of the governing equations and 
recovering the coefficients with higher accuracy due to averaging, especially for 
noisy data \citep{gure19}. The location of subdomains also matters. For instance, 
choosing the domains that don't overlap ensures linear independence of the rows 
of the $Q$ matrix. On the other hand, the location can also determine which physical 
effects are important. Poor sampling of the data can lead to models that are less 
general. For instance, viscous effects might not be picked up if boundary layers are 
not sampled \citep{gure21}. By using integration by parts on 
the amenable feature terms $\boldsymbol{q}^k_n$ we can transfer derivatives from
$\boldsymbol{u}$ to the weight function $\boldsymbol{w}$, thus greatly improving 
the robustness to noise. In this form weak SINDy can act as a low-pass
filter which averages over periodic high-frequency signals if $\Omega_k$ spans
periods of that frequency \citep{mess21}. 

Despite an increased robustness to noise, weak SINDy is still susceptible, as with 
any linear regression problem, to correlated terms in the feature library. This
presents a general issue for high-dimensional regression problems that despite
including some form of regularisation can still lead to models that are 
nonphysical. This has led to the inclusion of physical constraints that reduce 
the overall library size. For the weak form, the most general constraints are 
given by the SPIDER framework \citep{gure21, gold23} (described later), whereby 
the search is constrained to one of several group equivariant subspaces. 
While both pySINDy and SPIDER employ both weak formulation and sparse regression, 
the former requires specific domain knowledge to constrain the search space, 
unlike the latter. We present the application of both from a standpoint of 
an end user. 


\subsubsection{Mixed-integer sparse regression}

In this paper we make use of a recent 
addition to pySINDy termed mixed-integer optimiser for sparse regression (MIOSR) developed 
by \cite{bert23}. Sparsity is enforced in MIOSR through Specially Ordered Sets 
of Type 1 (SOS-1) restricting that only one variable in a set can take a nonzero
value.
By restricting the total number of allowable terms 
in the system, an answer is found which chooses the 
most suitable 
features in the linear regression problem. The solution is found using branch 
and bound algorithms which partition the feasible solution space into subsets of
problems and solve the linear relaxation of the problem at different nodes. This
process is continued until a solution with integer coefficients is found that
minimises the objective function and provides a solution to a combinatorially 
hard search problem \citep{bert23}. MIOSR relies on modern optimization 
solvers such as 
CPLEX\footnote{CPLEX, I.I.: V12. 8: User’s manual for cplex. Int. Bus. Mach. Corp. 46(53), 157 (2017)}
or GUROBI\footnote{Gurobi Optimization, LLC: Gurobi Optimizer Reference Manual (2021). https://www.gurobi.com} 
and we employ the latter here. Interested readers can refer to the original 
work by \cite{bert2016} for full details. 
When MIOSR is applied to all equations 
at once, the sparsity constraint $k$ is termed the group sparsity. Otherwise when 
it is applied to each equation individually it is referred to as target sparsity. 

There are three reasons we make use of this optimiser. First, MIOSR out-performs 
many of the other available optimisers on a series of benchmarking problems outlined
by \cite{kapt2023}. It is also the only optimiser that allows the inclusion of 
hard-constraints which we make use of to eliminate specific library terms. Finally,
as sparsity is enforced by limiting the maximum number of allowable terms appearing
in a given equation, terms with small coefficients can readily be identified. Given
that diffusive terms scale with $\sim 1/ \sqrt{R}$, the coefficients adorning these
terms will be small. The latter two points make the sequentially
thresholded least-squares approach as implemented in pySINDy unsuitable for this problem.

\subsubsection{Problem formulation in the pySINDy implementation}


In general when using pySINDy there are two approaches to library construction. 
The first is to create a general feature library of say, all product terms of 
the state variables and their derivatives up to an arbitrarily chosen order
of polynomial, and then if necessary constrain the resulting library. 
The second is to create a bespoke library which contains only the exact features required.
More details of how to create complex pySINDy libraries are shown in the
documentation\footnote{https://pysindy.readthedocs.io/en/latest/examples/15\_pysindy\_lectures/example.html\#Part-5:-How-to-build-complex-candidate-libraries}.
While the second approach is appealing, in practise we found this 
to be challenging to implement specifically in the weak form SINDy as it involves
taking many products of the input features and can still result in derivative terms
which are not intended. The simplest approach in pySINDy is to create a full general
product library and remove unwanted feature terms by constraining them. We return
to discuss the possible implications of this choice, particularly with respect to the
amount of computing memory consumed by pySINDy, in the Discussion section.

Herein, the pySINDy candidate library is constructed from the pressure $p$, the temperature $T$ 
and each component of the velocity, treated as a separate scalar feature object:
$u_x$ and $u_z$ for the 2D problems; $u_x$, $u_y$ and $u_z$ for the 3D problem.
To construct the custom library for 2D Rayleigh-B\'enard convection, we have 
included these four data objects $p(x,z)$, $T(x,z)$, $u_x(x,z)$ and $u_z(x,z)$
(4 terms), first-order partial derivatives (8 terms), second order partial
derivatives (16 terms, reduced to 12 applying chain rule), products
of the four data objects with the first-order partial derivatives (32 terms)
and products of the four data objects with the reduced second-order partial derivatives
(48 terms). A total of 104 possible library terms includes all the terms that
we know appear in the governing equations. Equivalently, the full candidate library
for the 3D problem contains 290 possible terms. Once this library has been generated 
by brute-force, pySINDy, scalar only in its operation, can then seek a relation 
for a predefined term by sparse regression and optimisation, which in the 
case of governing equations here is the time derivative for each feature object, 
e.g. $\partial_t p$. 

The library vector and vector of data objects together create a matrix of 
all coefficient possibilities, which we now artificially constrain using
knowledge about the problem in order to create
a library of physical terms which is more comparable to the manner in which
SPIDER generates a library. We take this artificial step in order to guide 
hyperparameter sweeping using an unconstrained library, as this is the
proper test of whether the machine learning can recover the governing
equations. We enforce $\partial_t p = 0$ by
explicitly including instruction not to seek a relation for this expression. 
We apply the knowledge that diffusion only appears in the governing
equations in the form of a Laplacian operator, thus reducing 60 second-order
terms to only 6. Furthermore, four of these are explicitly `switched off' for each
equation and the coefficients of the remaining terms in each equation are interlinked
(i.e. enforced to be equal to each other, in for example the inertial terms --- 
utilising knowledge that the equations are vectorial in nature).
For the artificial constraint of the library, we choose to apply the 
incompressibility constraint in the form
\begin{equation}
\partial_x u_x = -\partial_z u_z,
\end{equation}
removing 15 terms from the $u_x$ equation and 15 terms from the $u_z$ equation.
Elements of the full matrix of library terms are then interlinked,
i.e. enforced to be equal, so as to represent the coefficients of the
vector equations. The regression is thus forced to select exactly equal  
coefficient values for each component of every vector term (e.g., advection, 
diffusion and pressure term). We have used these artificial
choices, justified only by comparison to the known correct model, to construct 
and constrain the library only in order to guide the subsequent use
of unconstrained libraries, as we discuss below. Note that due to symmetry 
breaking induced by gravity there is no systematic {\it a priori} way either 
to choose the library terms or interlink the coefficients of equations describing 
different components of vectors (or higher-rank tensors) in SINDy, requiring a 
lot of {\it ad hoc} assumptions to be made.

When using weak SINDy there are several hyper-parameters and domain choices that must be made. The first is the volume $\Omega_k$ over which the integration is performed. In the 3D case we define the volume following \cite{rein20} 
\begin{equation}
    \Omega_k = \{(x,y,z,t): \vert x-x_k \vert \leq H_x, \vert y - y_k \vert \leq H_y, \vert z- z_k \vert \leq H_z,\vert t-t_k \vert \leq H_t \} 
\end{equation}
such that the volumes are centred around randomly selected points $(x_k, y_k, z_k, t_k)$ in the computational domain.
We interpolate from 384 Chebyshev grid space points vertically to a 
uniform 384-point grid
spacing, in order to match the uniform domain division applied in the 
formation of the library. This involves some loss of detail at the boundaries, which
we counter when necessary by increasing the number of uniformly spaced grid points
vertically.
A further issue is that positive-definite terms can become increasingly 
large with increasing $\Omega_k$, thus presenting scaling issues in the feature library, 
which degrade the quality of the fit. The total number of these subdomains $K$ must 
also be chosen and we have experimented with using between 300 and 1500, though in 
general more subdomains is more computationally expensive but yields better results. We explore the effect of varying 
these parameters in the Results section. We hold other parameters constant, typically
to their default values. For example, we use a NumPy random seed of 100 and we set the
order of the interpolating polynomial to be 6 in the initiation of the weak PDE library.

For hyper-parameters related to the optimiser, MIOSR only has a few that must be
selected. The first is $\alpha$ which relates to the strength of the $l_2$
regularisation. In general as the allowable sparsity is small we keep $\alpha$ small.
The second is the sparsity - either group or target. If we fit all equations at
once then the correct numbers of terms is 15 in total (group), corresponding to 
$5$ for $u_x$, $6$ for $u_z$, $0$ for $p$ and $4$ for $T$ (target = 5,6,0,4). 
Reducing this beyond these ``correct" values 
can lead to a loss of smaller contributing terms and can bring some understanding 
of the dominant balance of the equations. For example, one can search
for the Euler equations to test the hypothesis that the diffusive
terms will be lost first (with target = 3,4,0,2).
Increasing this beyond these values can `overfit' 
spurious extra terms, though this is not necessarily the case as MIOSR only has 
the requirement that the total number of terms be less than or equal to the total 
sparsity value. However, increasing 
the number of allowed terms does produce a more ill-posed regression problem.


\subsection{The SPIDER framework}

The SPIDER framework has been shown \citep{gure21} to be able to recover 
the complete mathematical model of a physical process (channel flow) - governing equations, 
constraints and boundary conditions. Unlike SINDy and its variants which all use an ad hoc
approach to constraining the search space, SPIDER is a complete framework that combines 1)
physical assumptions of smoothness, locality and symmetry for systematic construction of 
a collection of term libraries which define the search space; 2) a weak formulation of PDEs
for evaluating the contribution of different terms; and 3) model agnostic sparse regression
algorithm for inferring one or more parsimonious equivariant equations from each library.

For our purpose, SPIDER has some important differences to pySINDy. In particular, 
SPIDER is able to recover not only scalar equations, but also vector equations. 
Libraries in SPIDER are constructed by combining physical fields $\{ {\bf u}, p, 
\cdots \}$ with differential operators $\{\partial_t, \nabla \}$ using symmetry 
covariant operations such as tensor products and contractions. In this way, term libraries containing scalars, vectors, or even 
higher rank tensors can be constructed for model discovery. Symmetries further split 
libraries into irreducible representations via the construction of projection 
operators \citep{birdtracks} when possible.  

Following the example of the turbulent channel flow SPIDER analysis
\citep{gure21} we take the scalar pressure field $p$ and the vector velocity field 
${\bm u}$, add the scalar temperature field $T$ and combine them with 
differential operators. To account for the partial symmetry breaking associated with gravity, we add the unit vector $\bm{\hat{z}}$.
Using these objects, we can construct a  
feature library of possible candidate scalar terms
\begin{equation}
\begin{split}
\mathcal{L}_0\,=&\,
\{1,p,T,\nabla\cdot {\bm u},\,\partial_t p, p^2, p^3, \partial_t T, T^2, T^3, {\bm u}^2,
\,{\bm u}\cdot\nabla p, {\bm u}\cdot\nabla T, \nabla^2 p, \nabla^2 T, \\
&p\partial_t p,\,T\partial_t p,\,p\partial_t T,\,T\partial_t T,\,\partial_t^2 p,\,\partial_t^2 T,
\,p(\nabla\cdot{\bm u}),\,T(\nabla\cdot{\bm u}),\,{\bm u}^2p,\,{\bm u}^2T,
\,{\bm u}\cdot\partial_t {\bm u}
\} \\
\end{split}
\label{SPIDER-sca}
\end{equation}
and a second feature library of possible candidate vector terms
\begin{equation}
\begin{split}
\mathcal{L}_1\,=&\,
\{\bm{u},\partial_t\bm{u},\nabla p, \nabla T,\,p\bm{u},\,T\bm{u},(\bm{u}\cdot\nabla)\bm{u},
\nabla^2\bm{u},\,\partial_t^2\bm{u},\,u^2\bm{u},\,p^2\bm{u}, \\
&\,T^2\bm{u},\,\partial_t\nabla p,\,\partial_t\nabla T,\,p\nabla p,\,p\nabla T,\,T\nabla P,\,T\nabla T,\,\bm{u}(\nabla\cdot\bm{u}),\\
&\bm{u}\cdot(\nabla \bm{u}),\,\nabla(\nabla\cdot\bm{u}),
\,p\partial_t\bm{u},\,T\partial_t\bm{u},\,\bm{u}\partial_t p,\,\bm{u}\partial_t T,\,p\bm{\hat{z}},T\bm{\hat{z}}\}\\
\end{split}
\label{SPIDER-vec}
\end{equation}
which can be applied separately in searching for scalar or vector relations.
Employing Galilean invariance could reduce the size of the libraries further \citep{gure21}.
This should be the approach going forwards.

SPIDER yields a significantly smaller library; the 26-term scalar library forms the search space for any scalar fields, while
the 27-term vector library forms the search space for any vector fields. As noted
by \cite{gure21}, we emphasize that no domain knowledge specific to the problem, aside from
the symmetry (rotational and translational) and the choice of variables, has been used in
constructing these two libraries, in contrast to many other approaches. 
It should also be noted that libraries are modest in size compared to an equivalent 
104-term library constructed by the brute-force approach in pySINDy for the 2D problem.
Inclusion of physical symmetries and 
constraints can and does reduce these pySINDy libraries, as we have done above, but this is a
process that requires deep library understanding (exact knowledge of the ordering of
terms) and given the larger library size to be reduced, is naturally more susceptible 
to human error. 

The simplest scalar and vector relations describing pressure, temperature and velocity
data (the governing equations) can now be identified by performing a sparse regression
using these libraries $\mathcal{L}_0$ and $\mathcal{L}_1$.
Similarly to pySINDy, SPIDER uses the weak form of the PDEs following the approach in
\cite{gure19} in order to make the regression more robust. 
In this application, the numerical data contains only noise at the
level of numerical accuracy ($10^{-6}$), but even so, terms involving higher-order
derivatives, such as $\nabla^2 p$, $\nabla^2 T$ and $\nabla^2 \bm{u}$ are still 
sensitive to this noise. Specifically, the SPIDER framework multiplies each equation
by a smooth weight function $w_j({\bm x}, t)$ and then integrates it over a rectangular
spatiotemporal domain $\Omega_i$ of size $H_x \times H_z \times H_t$ (in 2D) or
$H_x \times H_y \times H_z \times H_t$ (in 3D). 

The derivatives are shifted from the data onto the
weight function whenever possible via integration by parts. The integrals are then
evaluated numerically using trapezoidal quadratures.  We use the same weight function 
forms raised to a power $\beta$ for the scalar and vector libraries as defined in 
\cite{gure21}. We similarly start from a value of $\beta = 8$ in our analysis and only
vary this if necessary, as this choice ensures that boundary terms vanish and maximises
the accuracy of the numerical quadrature along the uniformly gridded dimensions
\citep{gure19}. Given that for SPIDER, we maintain the non-uniform grid in the $z$ 
direction, increasing
$\beta$ further has no benefit as it only affects the errors of the quadrature, as 
illustrated in fig. 2 of \cite{gure21}.

In this work,
we vary the size of the subdomains $\Omega_i$, the number of subdomains and the value
of $\beta$ in order to recover the governing equations, as discussed in more detail below.

To summarise the method of \cite{gure21}, the repetition of this procedure 
for a number of domains $\Omega_i$ contained either within the full dataset or a 
specific region of interest (e.g. bulk flow or boundary region), constructs a 
feature matrix, which can be normalised to ensure the magnitudes of all the 
columns are comparable, the effect of which is to improve accuracy and 
robustness of the regression if necessary, although we don't find that necessary
for the recovery herein.
The resulting over-determined linear system is homogeneous and treats
all terms in the library on equal terms, in contrast to pySINDy. Complete details of the
procedure can be found in \cite{gure21}. Note that in the same result as that work, 
scalar terms are found that correspond to the incompressibility condition 
$\nabla\cdot\bm{u} = 0$ and its corollaries in the scalar library 
$p(\nabla\cdot\bm{u})$ and $T(\nabla\cdot\bm{u})$.
Now that the incompressibility condition has been identified, all the terms which involve
$\nabla\cdot\bm{u} = 0$ can be pruned from both the scalar and vector libraries, reducing
their complexity even further compared to brute-force approaches.
An iterative greedy algorithm \citep{gure21, gold24} then identifies multiple term relations from this pruned
library. Alongside this algorithm, a residual error is calculated; this error 
describes the weak form of the relations over all $K$ subdomains. Selecting a final 
relation from the comparison of reduced sequences formed by repeating the process and
dropping one term each time, is based on the choice between the simplicity
(i.e. number of terms $N$) and the accuracy quantified by the residuals, a choice 
which we explore in depth in the next section. Full details of the SPIDER framework have
been published elsewhere \citep{gold23,gure21} and we encourage the interested reader to review those 
works as well as several other relevant earlier studies \citep{gure19,rein20,rein21} on 
which this implementation of SPIDER is based\footnote{https://github.com/mgolden30/SPIDER}.

\section{Results}\label{results}

\subsection{pySINDy}

We have considered  the application of pySINDy only to 2D Rayleigh-B\'enard convection, owing to memory constraints;
it is our finding that pySINDy, for this method of application at least, quickly reaches 
the limits of the available RAM (196Gb) on the nodes of the supercomputing resources we had 
available to us. The high memory requirements of pySINDy for the PDEFIND method in application
to 2D and 3D data have been noted in the pySINDy documentation, with an example showing that a resolution of
$32\times 32 \times 32$ was the resolution possible for a 3D reaction-diffusion
system\footnote{https://pysindy.readthedocs.io/en/latest/examples/10\_PDEFIND\_examples/example.html}.
We discuss the possible reasons for this problematic consumption of
memory in the Discussion section. 

\subsubsection{Hyperparameter sweeping}

Without intuition about which hyperparameter values to choose, the natural start
is to perform a sweep across possible values. We set the location of the  spatiotemporal domains
according to the randomised selection of points detailed above and the division
of each domain ($x$, $z$ and $t$) extent by a constant for each dimension
($xdiv$, $zdiv$ and $tdiv$). In combination with
the number of spatiotemporal domains $K$, these constants are limited
at the smallest values (resulting in the largest spatiotemporal domains) by the memory
capacity available. At the largest values (resulting in the smallest spatiotemporal domains)
we limit our investigation to the maximum number of points in each direction for that
particular constant (theoretically true for the strong form only).

The aim of sweeping across these hyperparameters is to find reliable values
that can be used to recover governing equations from the unconstrained library.
We initially set out to recover governing equations from the constrained 
pySINDy libraries. The hope is that in future
work, any physical intuition that can be inferred from these values can be applied
in the search for unknown properties of the fluid (e.g. new sub-grid-scale
turbulence models). To assess the performance of the resulting models, we calculate 
the predicted root mean-squared errors over the $K$ subdomains for which the regression 
problem is formed.

Initially, we consider 2D Rayleigh-B\'enard convection at $R=10^6$. In the first three 
tests cases, the numerical results are interpolated from a Chebyshev grid to a uniform 
grid with 384 points in the vertical, with the horizontal resolution remaining the same. 
This is because pySINDy assumes a uniformly sampled spatiotemporal grid from the user in 
the numerical quadrature. Using a subset of the
high-resolution time data available, typically 400 points in the range $50 \leq t \leq 54$, 
because of the memory-demands, in the first test, we sample between 300 and 1500
spatiotemporal domains from the bulk of the flow - specifically in the region defined
on the ranges $2.60 \leq x \leq 3.60$ and $0.27 \leq z \leq 0.73$ (see box 1 in 
Figure \ref{figure1}). This corresponds to a region of the flow containing approximately
half of a quasistable periodic convective roll, including the plume and the centre 
of the roll. In the second test, we sample from the whole of the bulk, i.e. all $x$ and
$0.27 \leq z \leq 0.73$ (see box 2 in Figure \ref{figure1}). In the third test, we 
sample only from the lower boundary layer, following the hypothesis that this may be 
able to isolate the diffusive terms more easily, i.e. from the range $0 \leq x \leq 4$ 
and $0 \leq z \leq 0.13$ (see box 3 in Figure \ref{figure1}). The uniform grid of the 
first two cases has spacing set by the bulk of the flow, and thus under-resolves the 
boundary layers. In this case it is necessary to interpolate to a finer grid with 
spacing determined by the Chebshev grid at the boundary. The accuracy of numerical 
quadratures required by the weak form is substantially higher for uniform 
grids \citep{gure19}. Hence, in the fourth test, motivated by reducing the error
and improving the accuracy of the equations recovered, 
we interpolate the non-uniform vertical 384-point grid
defined by the Chebyshev polynomials onto a higher resolution uniform
grid of 3840 points and then select spatiotemporal domains from the full $x$ range
and from $0 \leq z \leq 0.026$.

For all these tests, we have explored the hyperparameter range of $x$, $z$ 
and $t$ divisions with a default of 300 spatiotemporal domains, which we have increased if necessary.
In the first test, we have used divisions in $x$ of 
($xdiv=8,10,12,14,16,18,20,22,24$), divisions in $z$ of ($zdiv=3,6,10,13,16,20$)
and divisions in $t$ of ($tdiv=8,10,12,16,20,24$). For the other tests, the set
of values used for $x$ and $z$ are scaled by the size of range sampled, such
that we maintain as close to the same size of spatiotemporal domains between
the tests. 

\begin{figure}
    \centering
    \includegraphics[width=1\linewidth]{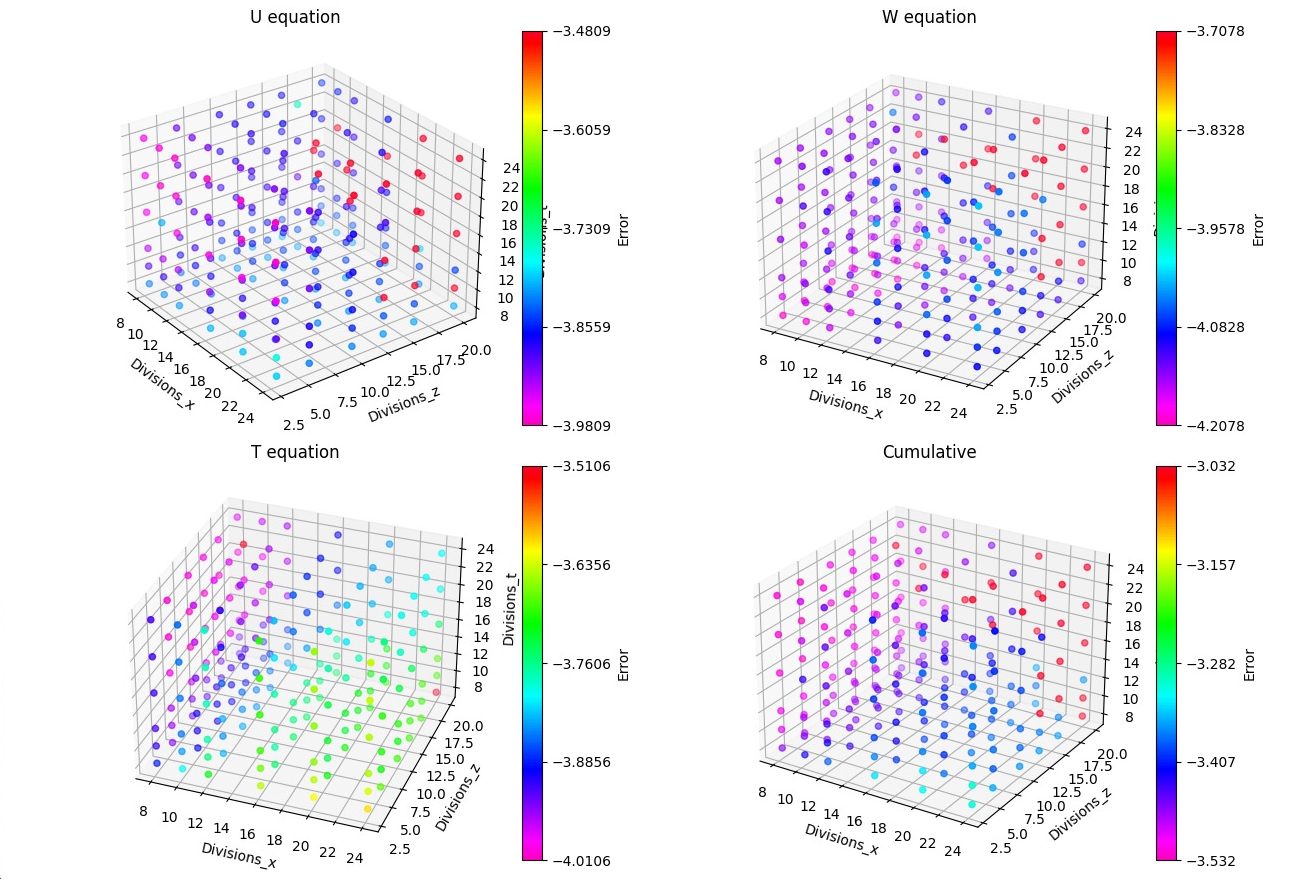}
    \caption{Results of the first pySINDy test at $R=10^6$ for 2D Rayleigh-B\'enard
    convection. Sampling spatiotemporal domains from a region of the bulk flow, the 
    figures show the effect of varying the size of the spatiotemporal domains on the
    resulting residual error (in weak form, over $K$ subdomains) of the governing equations recovered by the pySINDy 
    implementation. The error shown is the logarithm of the residual sum of squared errors.
    Specifically (a) visualises the residual error for the $u_x$ equation,
    (b) visualises the error for the $u_z$ equation, (c) visualises the error
    for the $T$ equation. Part (d) of the figure combines the residual errors
    for each equation together with the aim of finding a choice of $xdiv$, $zdiv$
    and $tdiv$ that corresponds to correct recovery of each governing equation
    independently; a choice of low $xdiv$, $zdiv$, and $tdiv$ appears to achieve 
    this aim.}
    \label{fig:bulk1}
\end{figure}

We illustrate the results of a hyperparameter sweeping test in Figure
\ref{fig:bulk1}. The results for each equation, for each combination of 
divisions is colour coded by the residual error of fit. Most published applications
of SINDy calculate some normalised measure of the coefficient error to 
assess model performance (or this choice is not discussed at all). We use 
this approach here, but it is a limited solution in our case as it relies 
on recovering only the correct equations (and knowing this beforehand
to constrain the libraries and be able to compare the result). 
The only other option available is to integrate the resulting PDEs and 
assess their performance, which is computationally infeasible for 
hyperparameter selection. This normalised cumulative coefficient residual
error is lowest in the region of small $xdiv$, $zdiv$ and $tdiv$,
specifically at $(xdiv,zdiv,tdiv) = (12,10,8)$.
Physically, these correspond to physically large spatial domains $(1024/12,384/10)$ with
the longest periods of time evolution $(400/8)$. We further find that equivalents of
these values are consistent between the first, second and third tests 
at $R=10^6$, implying that for this method, it is equally possible
to recover Navier-Stokes in both the bulk and at the boundary. 
The fourth test, sampling from the refined boundary layer, was inconclusive
suggesting that the region we sampled from, even with enhanced vertical resolution, 
was not large enough to recover both the viscous terms (which you would expect to 
dominate) and the advection terms. Further, any errors existing in the data will 
simply be interpolated to an increased grid resolution and thus still prevent 
successful recovery. We return to the question of data accuracy later. 
The governing equations recovered using the constrained library (where
we have constrained vector properties of the equations, as well as applied
the incompressibility constraint, in order to produce a more physical library)
typically lose the diffusive terms
first as you move away from this optimal combination, followed by the
addition of extra terms which are not found in the true equations.
This may be because weak SINDy introduces a smoothing effect if the domain 
is too large \citep{fung2024} or normalisation issues are introduced as 
terms become larger. For smaller integration domain sizes, there may be insufficient 
smoothing of errors. It is difficult to be conclusive as one thing we have
been unable to adjust when the MIOSR method is used within pySINDy is the
tolerance of the MIOSR method and we suggest this may be an avenue for
future investigation.

It is worth noting that each test required repeated applications of the
pySINDy implementation for each combination of $xdiv$, $zdiv$ and $tdiv$, corresponding
to different sized spatiotemporal domains. Each test took approximately 24 to 48 hours
on a single compute node of the ARC4 facility at the University of Leeds, 
at one number of $K$ spatiotemporal domains.
Future users may find it useful to note that we found that certain combinations, 
corresponding to the largest spatiotemporal domains, would attempt to use
more than the 196Gb of RAM available and would have to be run on the large
memory nodes with 768Gb RAM. 

\subsubsection{Increasing the Rayleigh number $R$}

\begin{figure}
    \centering
    \includegraphics[width=1\linewidth]{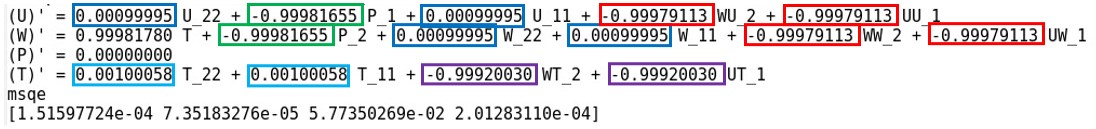}
    \caption{Results of constrained-library sparse regression 
    using the pySINDy implementation applied to
    simulation of 2D Rayeigh B\'enard convection at a Rayleigh number of
    $R=10^6$. Terms that are constrained by interlinking and have identical
    coefficients are indicated
    by colour boxes; gradient of pressure (green), advection (red and
    purple), diffusion (blue and yellow). Residual (root mean-squared) error
    is shown for each of the four equation on the bottom line. The third
    value (for the pressure equation) can be ignored. We
    choose to present this in pySINDy output format as an example of the
    results obtained by this analysis.}
    \label{1e6constrained}
\end{figure}

We now increase the value of $R$ (and hence the level of turbulence) and examine the ability of this pySINDy 
implementation to recover the Navier-Stokes equation governing velocity and the 
heat advection-diffusion equation governing temperature, continuing with the use of a constrained 
library. As demonstrated in the previous section, recovery of these
equations at $R=10^6$ was robust when the library was constrained
in the manner described above.
In Figure \ref{1e6constrained}, we show an example output of our pySINDy
analysis. Four equations are shown, all scalar, for the evolution of the
two components of velocity, the pressure and the temperature. There is no
pressure evolution equation, as expected from the constraints. The equivalent 
scalar equations following the notation of Equation \ref{RBeqns} are
\begin{equation}
\begin{split}
\frac{\upartial{u}}{\upartial{t}}\,=&
\,\textcolor{blue}{0.00099995}\,\frac{\upartial^2{u}}{\upartial{z^2}}
\,+\,\textcolor{green}{-0.99981655}\,\frac{\upartial{p}}{\upartial{x}}
\,+\,\textcolor{blue}{0.00099995}\,\frac{\upartial^2{u}}{\upartial{x^2}}\\
\,&+\,\textcolor{red}{-0.99979113}\,w\,\frac{\upartial{u}}{\upartial{z}}
\,+\,\textcolor{red}{-0.99979113}\,u\,\frac{\upartial{u}}{\upartial{x}},
\label{SINDy-ueqn}
\end{split}
\end{equation}
\begin{equation}
\begin{split}
\frac{\upartial{w}}{\upartial{t}}\,=&
\,0.99981780\,T\,
\,+\,\textcolor{green}{-0.99981655}\,\frac{\upartial{p}}{\upartial{z}}
\,+\,\textcolor{blue}{0.00099995}\,\frac{\upartial^2{w}}{\upartial{z^2}}\\
\,&+\,\textcolor{blue}{0.00099995}\,\frac{\upartial^2{w}}{\upartial{x^2}}
\,+\,\textcolor{red}{-0.99979113}\,w\,\frac{\upartial{w}}{\upartial{z}}
\,+\,\textcolor{red}{-0.99979113}\,u\,\frac{\upartial{w}}{\upartial{x}},
\label{SINDy-weqn}
\end{split}
\end{equation}
\begin{equation}
\frac{\upartial{p}}{\upartial{t}}\,=\,0,
\label{SINDy-peqn}
\end{equation}
\begin{equation}
\frac{\upartial{T}}{\upartial t}\,=
\,\textcolor{cyan}{0.00100058}\,\frac{\upartial^2{T}}{\upartial{z^2}}
\,+\,\textcolor{cyan}{0.00100058}\,\frac{\upartial^2{T}}{\upartial{x^2}}
\,+\,\textcolor{magenta}{-0.99920030}\,w\,\frac{\upartial{T}}{\upartial{z}}
\,+\,\textcolor{magenta}{-0.99929939}\,u\,\frac{\upartial{T}}{\upartial{x}}
\end{equation}
where the same colour coding has been used to highlight
values of coefficients that are enforced to be exactly identical to one
another - interlinked - by the constraint of the library.

A repeated set of tests at 
$R=10^8$ has shown similar ability to correctly recover the governing equations, again accurate for a range of hyperparameter spatiotemporal domain
values ($xdiv$, $zdiv$ and $tdiv$) around those optimally found for $R=10^6$.
At larger $R$, this robust result of the regression becomes less stable, 
meaning that even with a constrained library, at $R=10^{10}$ only the optimal 
parameters recover the governing equations and at $R=10^{12}$, only the 
Euler equation is recoverable. Again, this behaviour is 
consistent between tests 1, 2 and 3 and inconclusive for test 4. This is interesting
in of itself, as from physical intuition, natural expectation may expect
us to see different behaviour between these tests - we did not see that.
The difficulty of recovering governing equations at high $R$ may be due 
to the coefficient of
the diffusive term being a factor of $10^{6}$ smaller than the other coefficients
- equation discovery algorithms are known to struggle with a wide numerical
range of coefficient values. It should also be noted that the diffusive 
terms in the boundary layer are the smallest structures and therefore the
most difficult to resolve. In this instance we are inclined to conclude
that the tolerance of the MIOSR optimiser is set too high by default for
this case, but we also return to the effect of our chosen resolution 
on these machine learning methods later. Future investigation
is advised to understand the difference in equation recovery when sampling
from different regions of the flow, although we can say that applying a
post-computation increase in the number of points vertically by interpolation, simply
to provide a larger number of points to sample from for the selection of
spatiotemporal domains, does not improve the equation recovery performance;
the end-user of machine learning equation recovery techniques should be sure 
the numerical computations are well-resolved from the outset - a point we
return to in later discussion.

\subsubsection{Unconstrained library regression}

\begin{figure}
    \centering
    \includegraphics[width=1\linewidth]{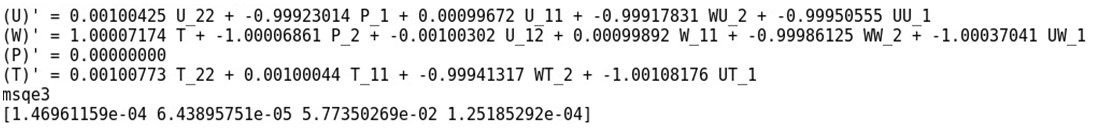}
    \caption{Results of unconstrained-library sparse regression 
    using the pySINDy implementation applied to
    simulation of 2D Rayeigh B\'enard convection at a Rayleigh number of
    $R=10^6$. Residual (root mean-squared) error
    is shown for each of the four equation on the bottom line. The third
    value (for the pressure equation) can be ignored.}
    \label{1e6unconstrained}
\end{figure}

It should be noted that with a constrained library, the correct form
of each term in the equations is selected in the regression. When using unconstrained
libraries, where all possible library terms are available, there are some
common swaps which appear in the recovered equations. In particular
for this incompressible problem, swaps related to the incompressibility condition
are common (though of course this is then simply a different correct way of writing the resulting equations --- though may lead to a different physical interpretation of the terms in the equations).

Noting that this occurs, with the optimally selected hyperparameters from the constrained
fitting, 
regression with the unconstrained library is able to recover the 
governing equations at the parameter
values which correspond to the minimal error, as shown in Figure 
\ref{1e6unconstrained} for $R=10^6$. A `swap' is notable in the $u_z$
(W') equation in the diffusive terms, where $\partial^2 u_z / \partial z^2$
has been replaced by $\partial^2 u_x / \partial x \partial z$ which is equivalent through
the incompressibility condition.
The residual errors are in fact slightly
smaller than in the constrained regression case, which shows that each
equation is being obtained separately, albeit from the same library.
Limited target sparsity was also enforced in the regression at an earlier result than
shown in order to see if simpler equations
(e.g. the Euler equation) are recovered first and this is indeed the
case. It is clear then that in order to recover the correct governing
equations with large libraries, if no physical intuition is decipherable from inspection of
the flow (a point we return to later) one must first optimise the values of the hyperparameters,
for example by using a brute-force repetitive computational test of
regression at all values to minimise the residual error.

It is also possible to alter the random
selection of the spatiotemporal domain locations
and statistically quantify some robustness
of the result.
We find that this process is effective at Rayleigh numbers of $10^6$
and $10^8$, in that with 100 variations of the random seed initialising
locations of selected spatiotemporal domains, the resulting equations
recovered are correct more than 75\% of the time for $R=10^6$ and more than
50\% of the time for $R=10^8$.
But even with tuned hyperparameters, the regression on
an unconstrained library is unsuccessful at $R=10^{10}$ and above.
More specifically, again it is possible to recover the simpler advection
terms robustly e.g. the Euler equation, but the recovery of the
diffusion terms, with their smaller coefficients,  has been
found to be very difficult - impossible at $R=10^{12}$. In future work, variation of the
order of the interpolating polynomial may be
able to improve on this poor results at high
$R$.

This is as far as we have been able to go with the pySINDy implementation.
Clearly for these kind of complex fluid flows, some physical intuition
is crucial for this method to constrain the candidate library of possible terms
for the sparse regression. Further, examining the flow may provide
some intuition about the optimal selection of hyperparameters 
(and we explore this and the limits imposed by resolution later). 
The generalised nature of
pySINDy places the onus on the end user to 
make appropriate choices in these areas before
investigation of more nuanced parameters like
the MIOSR tolerance and the order of the
integrating polynomial - which may in fact
have much greater bearings on the quality of
the resulting equation recovery. We 
discuss this further in the subsequent Discussion section.

\subsection{SPIDER}

\subsubsection{Governing Equations}

Inference of single-term relations in the SPIDER framework requires independent 
evaluation of all the terms in the corresponding library but does not involve 
regression \citep{gure21}. We verified that the incompressibility condition 
\eqref{RB-cont} is indeed recovered from the scalar library $\mathcal{L}_0$ for 
suitable choices of the sizes of integration domains.

\begin{figure}
\begin{center}
\subfigure[Momentum equation]{
\resizebox*{6cm}{!}%
{\includegraphics{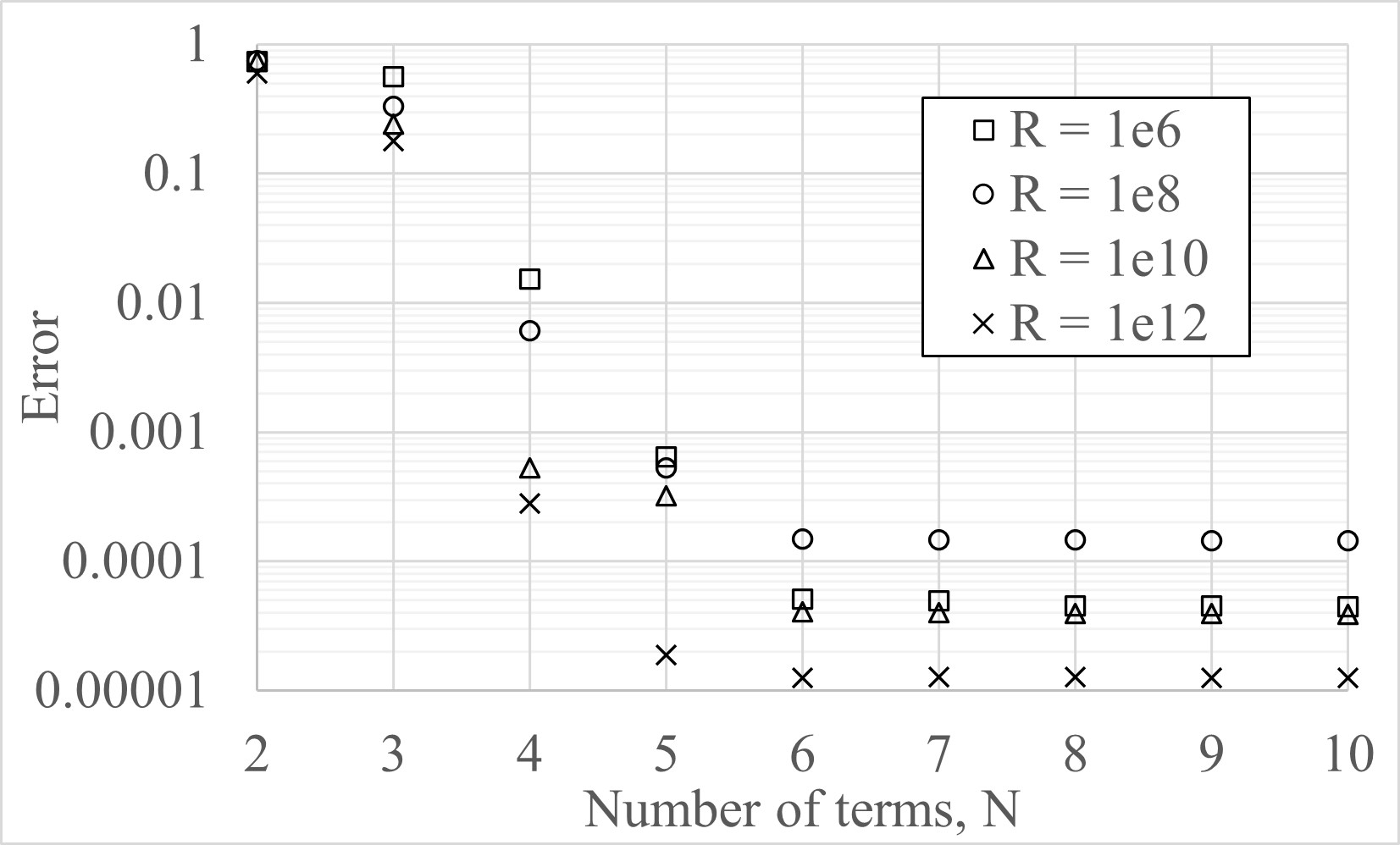}}}%
\subfigure[Heat advection-diffusion equation]{
\resizebox*{6cm}{!}%
{\includegraphics{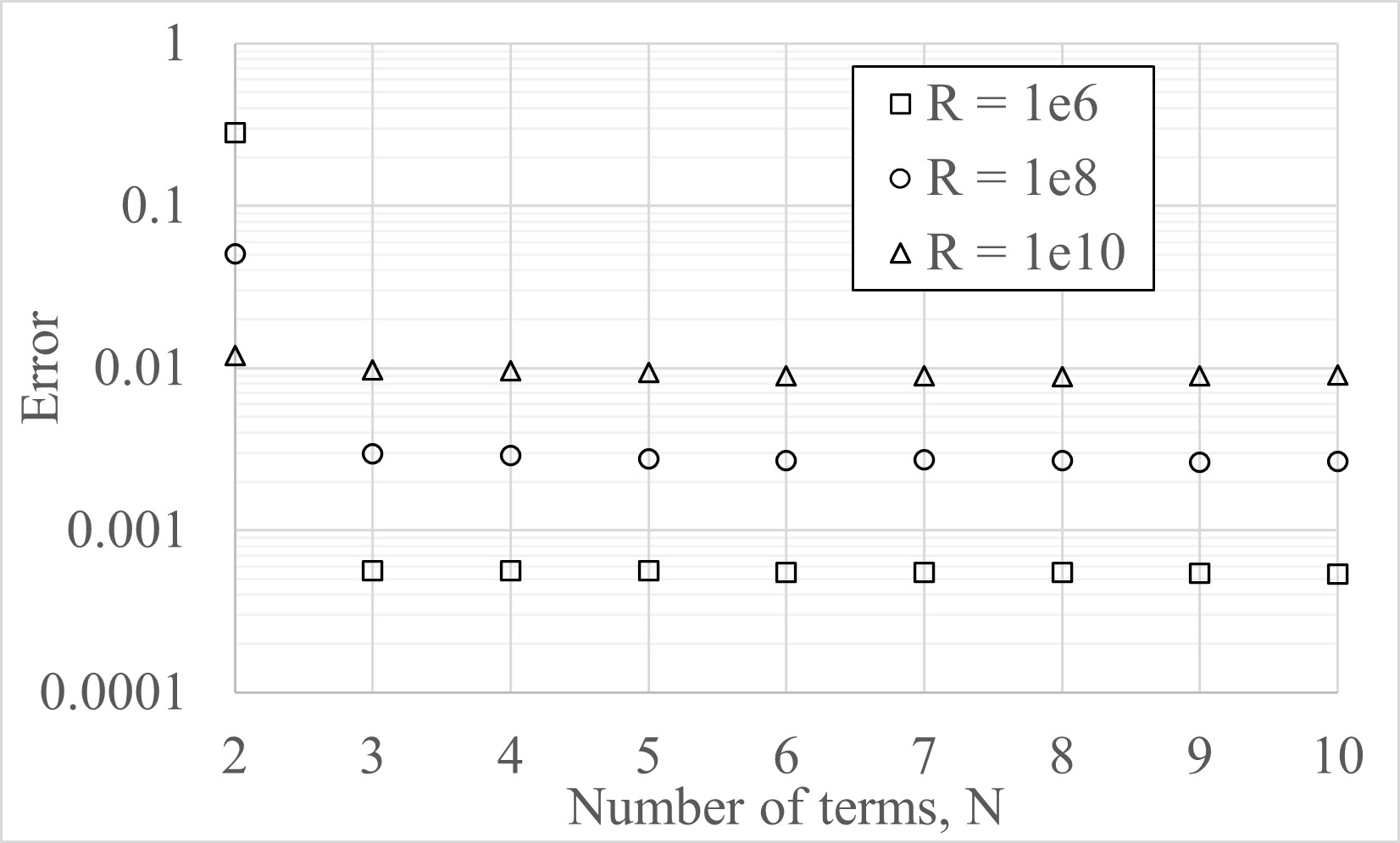}}}%
\caption{
Dependence of the residual on the number of terms $N$
retained in a recovered relation for the governing
(a) momentum equation and (b) heat advection-diffusion equation
for 2D Rayleigh-B\'enard convection.}
\label{figure-SPIDER-2DRBc}
\end{center}
\end{figure}

In Figure \ref{figure-SPIDER-2DRBc}(a) we show the results of the regression analysis performed
using the vector library $\mathcal{L}_1$ pruned for the incompressibility condition (see 
preceding section) in order to recover the momentum equation \eqref{RB-NSeqn}. 
We show the residual error computed as the number
of terms allowed in the library is reduced to only 1. The residual error for $N>10$ is not shown
as it follows the flat trend above $N=6$ for the momentum equation and $N=3$ for the heat advection-diffusion
equation. A natural choice of $N$, therefore, would either be 5 or 6 as this is the smallest number of terms
at which the error becomes more or less constant. In Table \ref{SPIDER-eqnresults} we show the details
of the momentum equation recovered by the SPIDER framework for $R=10^{10}$. It is not until four terms are 
selected that SPIDER recovers an evolution equation and this is reflected in the large reduction
in error at this $N$. It is reassuring to note that this is the Euler equation. With 5 terms, the 
SPIDER recovers the Navier-Stokes equation, with very close to the
correct coefficient for the diffusive term ($-10^{-5}\,\nabla^2\bm{u}$). 
Addition of a 6th term appears to reduce the error further. This extra term, indicated in Table
\ref{SPIDER-eqnresults} is spurious, but is likely to be fitting a real numerical error, 
which we discuss further in the next section.

From Figure \ref{figure-SPIDER-2DRBc}(a), one might naively deduce that the SPIDER framework
improves (i.e. produces smaller residuals) with increasing $R$. However, note that this masks the
adjustment of regression hyperparameters needed to obtain accurate recovery of the momentum equation,
which we required for the largest $R$. More specifically, at $R=10^6$, we used 1024
spatiotemporal subdomains, $\beta=12$ and $\Omega_i = 128$. At $R=10^8$, we can reduce $\Omega_i$
to 64. For $R=10^{10}$ we were able to recover governing equation with only 128 subdomains, 
$\beta=4$ and again $\Omega_i = 64$. At $R=10^{12}$, the best recovery parameters we found of
$\Omega_i = 128$, $\beta=4$ and 1024 subdomains were able to recover the Euler equation with 4
terms, but the SPIDER framework consistently selected a $\partial_t\nabla p$ term
before the diffusive term. Given that the coefficient of the diffusive term should be $10^{-6}$,
perhaps it is no surprise that any method struggles to reproduce such coefficients close to the numerical
precision of the simulation. Or, it is possible that SPIDER is telling
us something about the quality of the data - we explore this question
in the Discussion section.
All the same, the recovery of the Euler equation alone by the SPIDER
framework for $R=10^{12}$ and the robust recovery of Euler and Navier-Stokes equations for $R=10^{10}$
is a remarkable performance improvement over pySINDy, at least for this problem and in the manner we 
have applied both equation discovery methods.

The larger value of $\Omega_i$ required for the lowest $R$ laminar flow is not surprising, since the structures
of the laminar flow are large convective rolls and this larger $\Omega_i$ will sample larger
regions of the flow in both space and time. Whilst we have presented results with varying $\beta$, 
in reality the effect of varying this parameter with Fourier-Chebyshev data is small compared to
the effect of variation in the number of subdomains and very small compared to the effect of 
variation of subdomain sizes. Since $\beta$ would only be expected to improve the regression in 
the $x$ direction, then this is no surprise. 

\begin{table}
\tbl{Vector equations recovered by the SPIDER framework from 2D simulation of
Rayleigh-B\'enard convection at a Rayleigh number $R=10^{10}$,
indicative of the same term order recovery with different coefficients for
smaller Rayleigh number$^{\rm a}$.}
{\begin{tabular}{@{}lccccccc}\toprule
N & Error & \multicolumn{6}{c}{Coefficients of terms$^{\rm b}$} \\
 & & $\partial_t\bm{u}$ 
   & $(\bm{u}\cdot\nabla)\,\bm{u}$ 
   & $\nabla p$ 
   & $T\,\bm{\hat{z}}$ 
   & $\nabla^2{\bm{u}}$
   & $\partial_t\nabla p$ \\
\colrule
2 & 0.740 & - & - & 0.579$^{\rm c}$ & -1$^{\rm c}$ & - & - \\
3 & 0.242 & - & 0.882 & 0.960 & -1 & - & -\\
4 & $5.261\times10^{-4}$ & 1 & 0.999 & 0.999 & -0.999 & - & - \\
5 & $3.219\times10^{-4}$ & 0.999 & 1 & 1 & -0.999 & $-1.023\times10^{-5}$ & - \\
6 & $4.048\times10^{-5}$ & 1 & 1 & 1 & -0.999 & $-1.006\times10^{-5}$ & $-5.112\times10^{-4}$ \\
\botrule
\end{tabular}}
\tabnote{$^{\rm a}$Whilst recovery of the 4-term Euler equation is consistent across
all $R$, this table is not indicative of recovering the Navier-Stokes equation for
$R=10^{12}$. Please see the text for further explanation.}
\tabnote{$^{\rm b}$The sign of the terms here is shown as if
everything was moved to the left-hand side of the momentum equation
i.e. $\partial\bm{u}/\partial t + ... = 0$.}
\tabnote{$^{\rm c}$The results produced by the SPIDER framework do
indicate a difference between exact integers and decimals.}\label{SPIDER-eqnresults}
\end{table}

In Figure \ref{figure-SPIDER-2DRBc}(b), we show the results of the regression analysis 
performed using the scalar library $\mathcal{L}_0$ pruned for the incompressibilty condition 
and also for time derivatives of $p$ (akin to deselecting the 
$\partial p / \partial t$ equation in the pySINDy libraries) in order to recover the heat advection-diffusion equation  \eqref{RB-Teqn}.
The residual error goes constant from $N=3$ terms and it is reassuring to see that the 3-term
equation recovered by SPIDER for Rayleigh numbers $10^6$, $10^8$ and $10^{10}$ is the correct
heat advection-diffusion equation, as shown in Table \ref{SPIDER-heateqnresults}. In this
analysis, we used 1024 subdomains and $\beta=8$ for all $R$. We also set $\Omega_{x,z}=64$
and $\Omega_t=128$ for both $R=10^6$ and $R=10^8$. For $R=10^{10}$ we found it necessary to reduce
the size of the subdomains to $\Omega_i=48$, indicative of the smaller scale more turbulent
structure and thinner boundary layer at this $R$. It is noteworthy that despite wide 
investigation, including converting the vertical resolution from 384 points in 
a Chebyshev grid to 1024 and 2048 points on a Fourier grid and then experimenting
with varying $\Omega_i$ and $\beta$, we have only been able to recover a heat
advection equation at $N=2$ (and even then only with a reduced scalar library that removes 
other time derivative terms than $\partial T/\partial t$), but never the heat 
advection-diffusion equation for larger numbers of terms; the residual error of this
is not shown in Figure \ref{figure-SPIDER-2DRBc} as it is ~0.01 for $N\geq2$.
Similarly to the vector equation, where more coefficients 
are tied together across
the vector components, thus possibly making the regression more robust, the identification of the 
diffusive term with a coefficient close to numerical accuracy is clearly at
(and beyond for $R=10^{12}$) the limit of the algorithms for this data. We
also limited the analysis to the boundary layer only, but were still unable
to recover the full equation. One might also note that the flow is also highly turbulent at this $R$ and
that this incoherence and short correlation lengths and times might pose difficulty for the ML 
methods, but as we shall demonstrate with three-dimensional simulations, 
this does not seem to affect the capability of the SPIDER framework to recover the governing equations.
In fact, in three dimensions, the accuracy of the recovery improves for both the Navier-Stokes and heat 
advection-diffusion equations at $R$ values which result in a turbulent flow, as we discuss next.

\begin{table}
\tbl{Scalar equations recovered by the SPIDER framework from 2D simulation of
Rayleigh-B\'enard convection at a Rayleigh number $R=10^{10}$,
indicative of the same term order recovery with different coefficients for
smaller Rayleigh number.}
{\begin{tabular}{@{}lccccc}\toprule
N & Error & \multicolumn{3}{c}{Coefficients of terms$^{\rm a}$} \\
 & & $\partial_t T$ 
   & $(\bm{u}\cdot\nabla)T$
   & $\nabla^2 T$
   & $T^2$ \\
\colrule
2 & $1.191\times10^{-2}$ & 0.997 & 1 & - & -\\
3 & $9.694\times10^{-3}$ & 0.997 & 1 & $-1.009\times10^{-5}$ & -\\
4$^{\rm b}$ & $9.613\times10^{-3}$ & 0.997 & 1 & $9.814\times10^{-6}$ & $1.598\times10^{-3}$\\
\botrule
\end{tabular}}
\tabnote{$^{\rm a}$The sign of the terms here is shown as if
everything was moved to the left-hand side of the momentum equation
i.e. $\partial T/\partial t + ... = 0$.}
\tabnote{$^{\rm b}$At $N=4$, an anti-diffusive term is recovered,
along with a spurious $T^2$ term, demonstrating that care has to be
taken to physically interpret the recovered results for reasonable
solution to the observed problem.}
\label{SPIDER-heateqnresults}
\end{table}

\begin{figure}
\begin{center}
\subfigure[Momentum equation]{
\resizebox*{6cm}{!}%
{\includegraphics{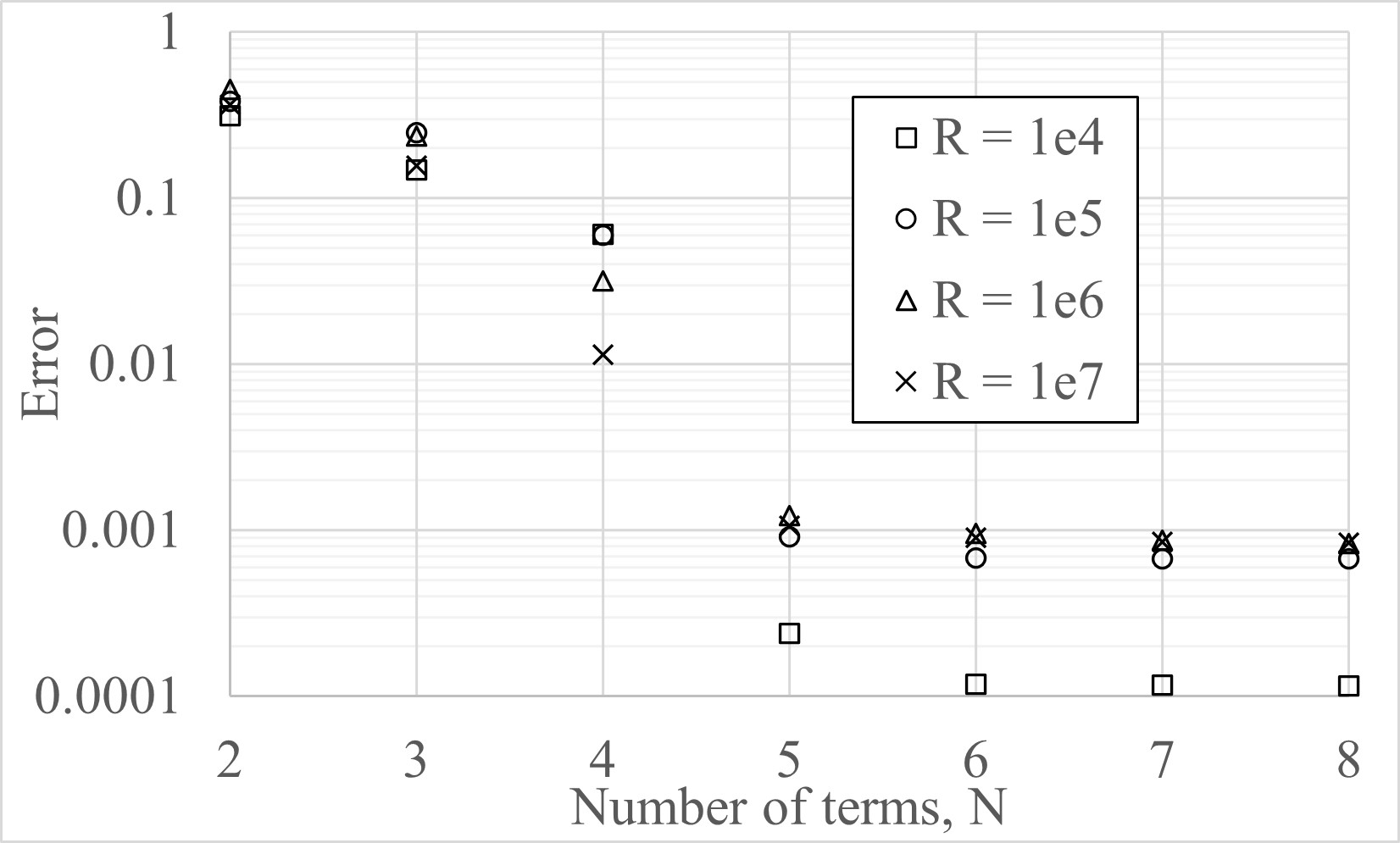}}}%
\subfigure[Heat advection-diffusion equation]{
\resizebox*{6cm}{!}%
{\includegraphics{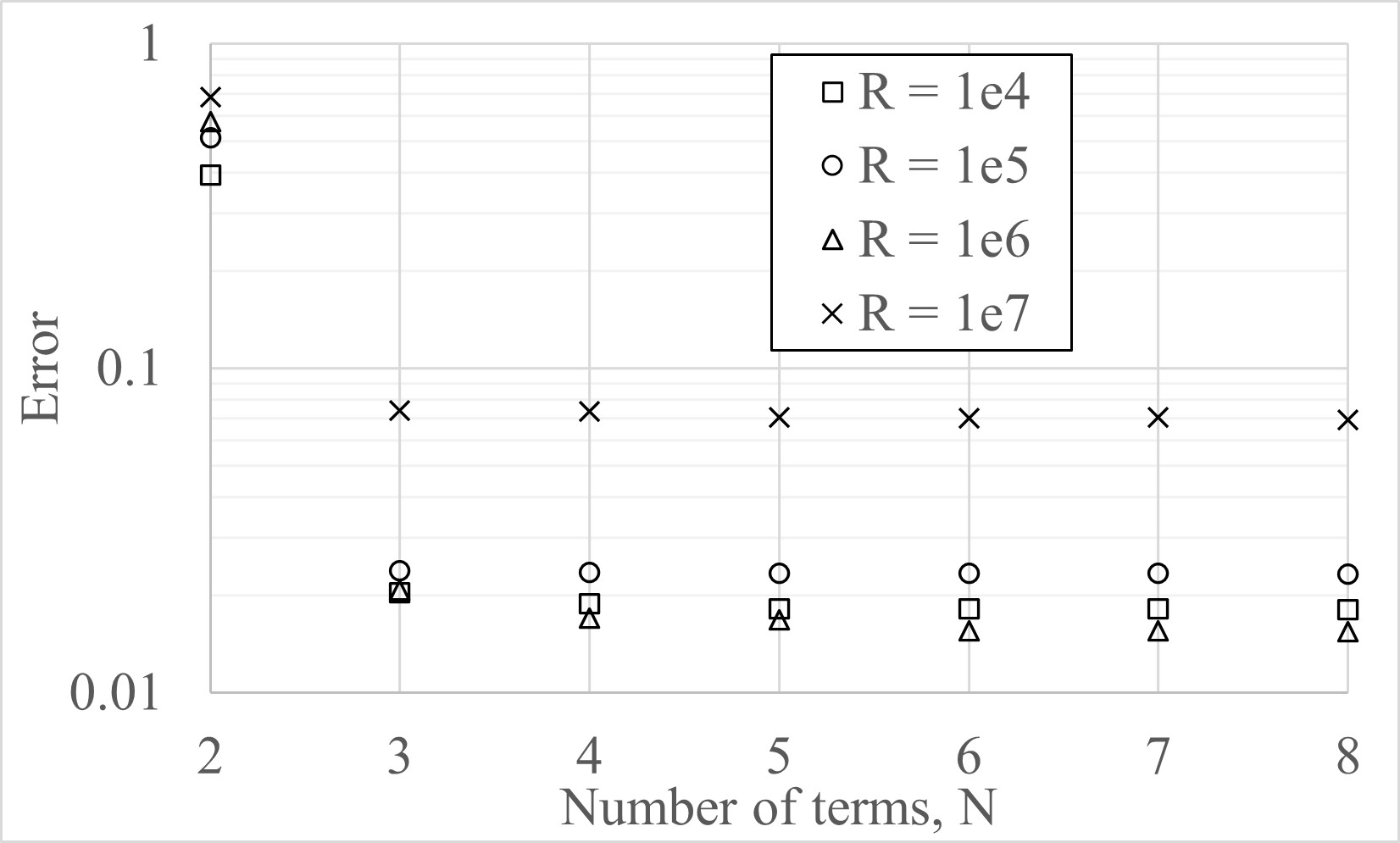}}}%
\caption{
Dependence of the residual on the number of terms $N$
retained in a recovered relation for the governing
(a) momentum equation and (b) heat advection-diffusion equation
for 3D Rayleigh-B\'enard convection.}
\label{figure-SPIDER-3DRBc}
\end{center}
\end{figure}

Figure \ref{figure-SPIDER-3DRBc}(a) presents the results of the regression analysis 
performed using the vector library $\mathcal{L}_1$ in order to recover the momentum
evolution equation in three dimensions. In contrast to the two-dimensional analysis, this time the 4-term Euler equation
and 5-term Navier-Stokes equation are recovered before any other expression for the whole
range of $R$ examined from $10^4$ to $10^7$, which again covers the laminar, transitional
and turbulent regimes. The regression parameters are shown in Table 
\ref{SPIDER-3DRBc-params}. It was possible during this analysis to keep the regression
parameters fairly consistent, so as to be able to examine the effect of $R$ on the 
residual error and it can be seen from Figure \ref{figure-SPIDER-3DRBc}(a) that a smaller
residual error corresponds to a smaller $R$, or in other words a more laminar flow pattern with larger structures,
which is more coherent and correlated in both space and time, is easier to recover by
these methods.
This is perhaps not a surprising result - that such an  algorithm would have 
less difficulty with a smoother, less turbulent flow - but interesting to note 
and reassuring that physical intuition is aligned with the SPIDER framework.

\begin{table}
\tbl{Parameters of the vector regression analysis used for the application
of the SPIDER framework to simulations of 3D Rayleigh-B\'enard convection.}
{\begin{tabular}{@{}lcccccc}\toprule
$R$ & $N_{windows}$ & $\beta$ & $H_x$ & $H_y$ & $H_z$ & $H_t$\\
\colrule
$10^4$ & 512 & 8 & 32 & 32 & 64 & 64 \\
$10^5$ & 512 & 8 & 32 & 32 & 64 & 64 \\
$10^6$ & 256 & 4 & 32 & 32 & 32 & 32 \\
$10^7$ & 256 & 4 & 32 & 32 & 32 & 32 \\
\botrule
\end{tabular}}
\label{SPIDER-3DRBc-params}
\end{table}

Figure \ref{figure-SPIDER-3DRBc}(b) presents the results of the regression analysis 
performed using the scalar library $\mathcal{L}_0$ in order to recover the heat
advection-diffusion equation for three dimensional convection. In an improvement over the analysis of the 2D 
Rayleigh-B\'enard convection data, here the 3-term heat advection-diffusion equation
is robustly recovered at all $R$. Regression parameters are held constant at
1024 subdomains, $\beta=4$ and $\Omega_i=32$ for $R=10^4$, $R=10^5$ and $R=10^6$ and
the residual error is consistent across all these values, albeit higher than for two-dimensional
convection. In order to recover the correct equation and coefficients for the 
simulation of 3D convection at $R=10^7$, 
2048 subdomains, a value of $\beta=8$ and $\Omega_i=[16,16,32,32]$ were required and 
even then the residual error is far from ideal. Turbulence in the flow does clearly
make the regression analysis more difficult, but unsurprisingly there is also a 
distinct cut off in the ability of machine learning methods as the numerical accuracy 
of the data is reached. 

\begin{figure}
\begin{center}
\subfigure[Momentum equation]{
\resizebox*{6cm}{!}%
{\includegraphics{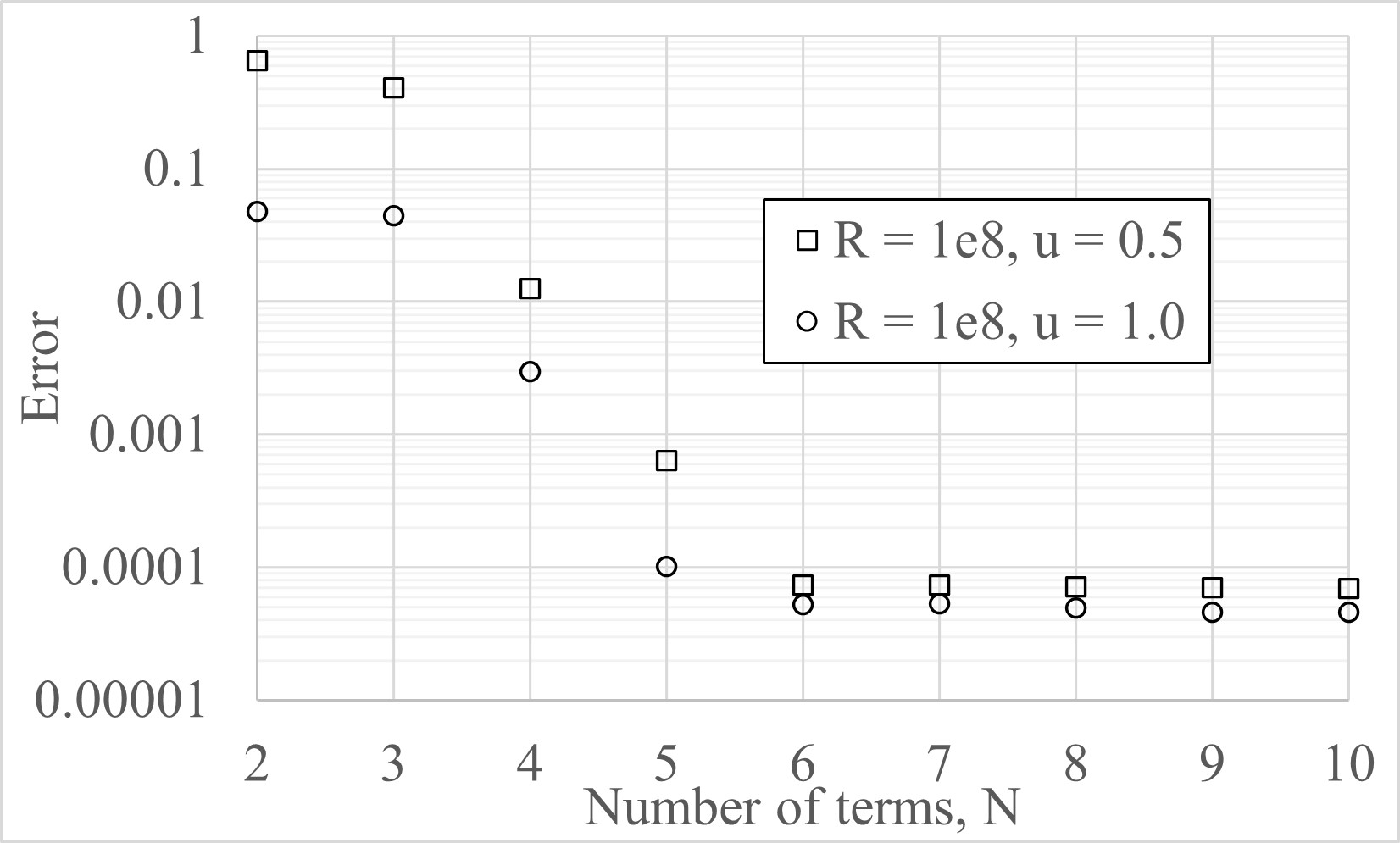}}}%
\subfigure[Heat advection-diffusion equation]{
\resizebox*{6cm}{!}%
{\includegraphics{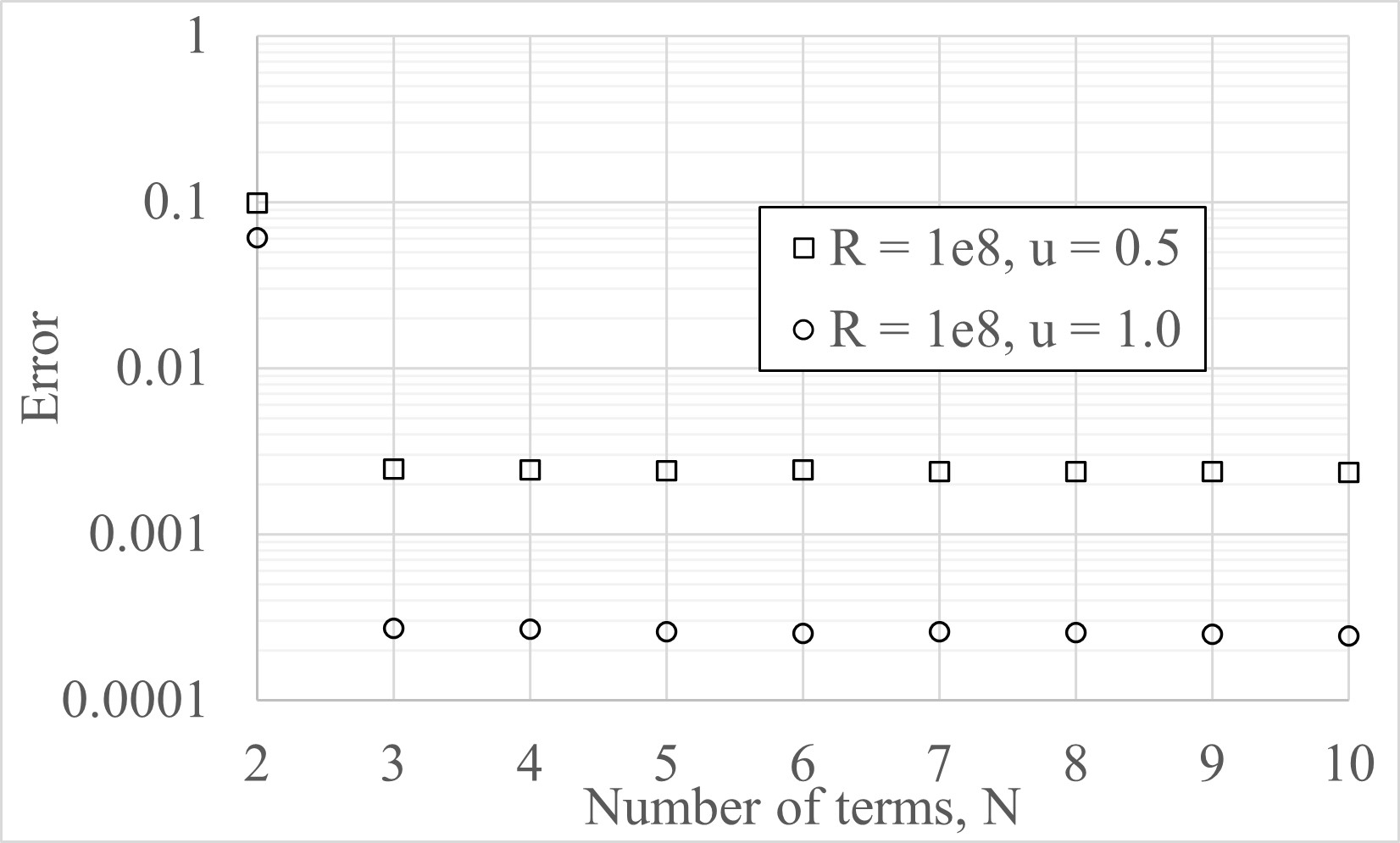}}}%
\caption{
Dependence of the residual on the number of terms $N$
retained in a recovered relation for the governing
(a) momentum equation and (b) heat advection-diffusion equation
for 2D planar convective Couette flow.}
\label{figure-SPIDER-2Dcou}
\end{center}
\end{figure}

We show the results of the regression analysis performed using the same
libraries applied to the simulated data of 2D planar convective Couette flow in
Figure \ref{figure-SPIDER-2Dcou}. Similarly to the previous two cases, the SPIDER
framework is again able to robustly recover the 4-term Euler equation and the 5-term
Navier-Stokes equation for the momentum evolution and the 3-term heat 
advection-diffusion equation before including other terms. 
These analyses have both used 1024 subdomains. 
For the vector analysis $\beta=12$ and $\Omega_i = [64,64,64]$ produced better
residuals and for the scalar analysis $\beta=8$ and $\Omega_i = [64,64,128]$
produced better residuals. The velocity of the top and bottom planes did not
seem to affect the regression analysis in the recovery of governing equations.

\citet{gure21} noted that for noiseless data, as we have here, the asymptotic
value of the residual is determined by the discretization of the data, which would
imply acceptable residual errors on the order $10^{-3}$, as in general is consistent with what is observed
here.

\subsubsection{Boundary conditions}

The recovery of boundary conditions using the SPIDER framework requires a different approach to
the libraries, because the rotational symmetry assumption is broken in 2D and partially broken 
in 3D - the problem is only invariant with respect to rotations about the normal $\hat{\bm{n}}$ to the
boundary. To overcome this problem, we follow the approach of \cite{gure21} and include
$\hat{\bm{n}}$ in constructing a library $\mathcal{L}_{\bm{u}}$ of terms that transform as vectors, whilst excluding time
derivatives. 
Similarly, to account for the symmetry breaking associated with the wall motion, we include the unit vector $\hat{\bm{x}}$ in that direction.
Finally, we exclude the scalar fields $p$ and $T$ which yields
\begin{equation}
\mathcal{L}_{\bm{u}}\,=\,
\{\hat{\bm{n}},\,\bm{u},\,
\nabla(\bm{u}\cdot\hat{\bm{n}}),\,
(\hat{\bm{n}}\cdot\nabla)\bm{u},\,\hat{\bm{x}}
\},
\label{SPIDER-bclib}
\end{equation}
where $\hat{\bm{n}}=\pm\hat{\bm{z}}$, depending on whether the top or the bottom boundary is considered.
We then split this library into the irredicible representations, an invariant and a covariant, with respect to rotations about $\hat{\bm{n}}$ (corresponding to the
normal and tangential components, respectively), by applying the projection operators $P_\bot\,=\,\hat{\bm{n}}\hat{\bm{n}}$ and
$P_\parallel\,=\mathbbm{1} - \hat{\bm{n}}\hat{\bm{n}}$. This yields the following separate libraries
for the velocity
\begin{equation}
\begin{split}
\mathcal{L}_\parallel\,=&\,
\{P_\parallel\,\bm{u},\,
P_\parallel\nabla(\bm{u}\cdot\hat{\bm{n}}),\,
P_\parallel(\hat{\bm{n}}\cdot\nabla)\bm{u},\,\hat{\bm{x}}
\},\\
\mathcal{L}_\bot\,=&\,
\{\hat{\bm{n}}\cdot\bm{u},\,
1,\,
\hat{\bm{n}}\cdot\nabla(\bm{u}\cdot\hat{\bm{n}}),\,
\hat{\bm{n}}\cdot(\hat{\bm{n}}\cdot\nabla)\bm{u}
\}.\\
\end{split}
\label{SPIDER-bclib-u}
\end{equation}
In 2D, these libraries
reduce to the components of each field and the derivatives of the components either parallel or
perpendicular to the boundary. 
A scalar library $\mathcal{L}_T$ for the temperature field $T$ can be constructed in a similar fashion
\begin{equation}
\mathcal{L}_T\,=\,
\{1,\,T,\,
(\hat{\bm{n}}\cdot\nabla)T,\,
(\hat{\bm{x}}\cdot\nabla)T
\}.
\label{SPIDER-bclib-T}
\end{equation}
Only data on the boundary is sampled, effectively selecting a
physical profile (for 2D data) or slice (for 3D data) which has variation with time. Again, a number
of spatiotemporal subdomains are sampled, typically 256, of a specific size, typically
$\Omega_i=64$. $\beta=4$ is used. As integration in the wall-normal
direction cannot be performed for these samples, a finite differencing of the data has to be performed on the full data
and stored, before the original data arrays and the arrays of their derivatives are reduced to obtain
a profile or slice. In application to the Rayleigh-B\'enard convection problems and specifically 
only the solid boundaries at $z=0$ and $z=1$, the correct temperature and stress-free velocity
boundary conditions are recovered {\it for all $R$ in both 2D and 3D}. In application to the 2D planar convective Couette flow problem,
again the correct temperature and velocity no-slip (fixed velocity of 0.5 or 1.0) boundary
conditions are recovered. This demonstrates another advantage of SPIDER in being able to
recover boundary conditions, as well as governing equations and constraints.

\section{Discussion}\label{discuss}

It has been difficult to find other works which are as revealing about finding
the hyperparameters which work well for pySINDy as we have been herein. This is not
a simple process and in order to perform the methodical hyperparameter sweeps 
we have performed above, requires access to computing resources with
significant amounts of memory. Further,
assessing the results of these sweeps is also tricky. 

The key difference between pySINDy and SPIDER is physical constraints. In the pySINDy 
implementation, the brute force approach
combines everything together and the end user is left to use the resulting large library or manually
reduce this library in a non-simple manner. The linking of terms across the regression matrix to
make coefficients of vectors adds another layer of complexity. 
Extending pySINDy to libraries containing higher-rank tensors while ensuring equivariance is prohibitively difficult.
SPIDER crucially relies on the
physics to define the library - more specifically the irreducible representations of the symmetry
group describing the physical problem, defined by the data and differential operators such as $\partial_t$ (or $\partial_t+\bm{u}\cdot\nabla$ in the Galilean-invariant case)
and $\nabla$. This is really the crucial difference between pySINDy and SPIDER that is shown clearly
in the size of the respective libraries and that makes SPIDER more intuitive to use, at least for fluid
dynamicists such as ourselves.

It is also interesting to reasonably hypothesize that the memory problem we have observed when
using pySINDy, even in 2D, may be related to the size of the library. Even with terms `disabled', 
the column for that term still exists in the matrix and therefore consumes computing memory in
the regression calculation. A way to overcome this memory problem in future work may therefore
be to only use very custom libraries which mimic the same physical approach as that used in
our SPIDER framework. We emphasize that this is not currently a simple task to consider, especially
when further interlinking of terms in this library is then required to mimic vector (or even higher-rank) equations.
But, we also emphasize, this is our suggested route to overcoming the memory problem we have
observed and that is opaquely referred to in pySINDy documentation (see earlier in this work
for specific details).

It is interesting to consider how these methods may be used to discover unknown, rather than 
known, governing equations, constraints and boundary conditions
as we have detailed here. In fact, we intend to apply these same methods to
the search for turbulent closure models such as \citet{gar10} using our insight from this work
and that of other authors \citep{jak24}. In particular, for pySINDy, it is clear that parameter
sweeps of the regression hyperparameters do result in finding the best combination of these
parameters for the analysis, which can then be used as, for example, $R$ is increased, as we did
herein, or in future may be used to construct the default position for the uninformed 
turbulence closure
analysis. We expect such similar parameter sweeps may be easily done in the future with SPIDER as and
when the python library for SPIDER becomes available. This would also be aided by the ability
to fit higher order tensors. In combination with careful library selection
of suitable terms, e.g. triple correlation velocity terms, this regression tuning for a particular
flow will be key to uninformed discovery with either machine learning method used here.
Such parameter sweeping with SPIDER as we did with pySINDy, when possible, may fill in the gaps 
and/or extend the ability to fine tune the regression parameters and solve the open 
questions/limits of SPIDER above. For any turbulence closure analysis, we remind the interested
reader that there is also the question of separating the mean and fluctuating components of
the flow, a fuller discussion of which can be found in \cite{jak24}. 

It is notable that only a certain set of hyperparameters were varied for the pySINDy approach.
Throughout the pySINDy analyses, we held the order of the integrating polynomial constant at 6,
in order to make the exploration tractable with the resources available as much as anything else.
With SPIDER, it was an easier proposition to alter the equivalent $\beta$ parameter and we did so
as detailed above. Future pySINDy work should consider the variation of the order of the integrating
polynomial if the four-dimensional hyperparameter exploration is possible.

We have observed that both pySINDy and SPIDER are capable of recovering small spurious terms
in the governing equations, likely to be fitting real error.
In our experience, this error is either some small numerical error in the fields
(common in low order integration/reconstruction schemes) or it is possible that the weak 
form subdomains are not matched to the length/time scale of the dynamics. We have observed that
tuning of the size of the subdomains is key to obtaining the ``correct" result and requires
investigation for each specific instance of $R$, making it difficult to generalise the method
for this flow problem and only possible for a specific value of $R$. This will be useful 
to recall as our investigation moves to recovery of unknown equations, specifically the 
recovery of existing and new turbulence closures. Thankfully we are not dealing with a large
degree of numerical noise or incomplete data. The machine learning process we detail
may be possible with such data, but we leave that to future work which could be easily
investigated from the Dedalus, pySINDy and SPIDER scripts we publish alongside this paper.

It has been emphasized previously \citep{gure21} that a proper 
non-dimensionalisation of the data is required. The non-dimensionalisation of the model, 
chosen here for numerical convenience more than anything else
(see Section \ref{models}), seems to have been effective
for both the machine learning analyses. It is possible that a better 
non-dimensionalisation, or further
a recasting into different coordinates (as noted is particularly effective for certain problems
in pySINDy), may improve the performance we have observed here.
An exploration of different non-dimensionalisations and such recasting
is beyond the scope of this work.

\subsection{Using domain knowledge to choose subdomain sizes - correlation scales}

It should be possible to avoid the kind of hyperparameter sweeping that we have done here,
as analysis of the flow can inform the initial choice of the size of the subdomains.
In particular, calculation of the correlation length and time of the flow from the data may give some good \textit{a priori}
indication of the optimal spatiotemporal domain. That said, in the SPIDER analysis,
the box sizes were held constant in terms of the number of grid points - 
{\it which crucially allowed the boxes to be physically smaller in the vertical direction
closer to the boundaries as an effect of the non-uniform spacing of the grid}. It is possible
that this approach is key to the improved performance we have obtained with SPIDER.

\begin{table}
\tbl{Correlation lengthscales and timescales for the simulations of 2D
Rayleign B\'enard convection.}
{\begin{tabular}{@{}lcccc}\toprule
$R$ & $x_{corr}$ & $z_{corr}$ & $\ell_{corr} = (x_{corr}^2 + z_{corr}^2)^{0.5}$ & $t_{corr}$\\
\colrule
$10^6$ & 0.49 & 0.23 & 0.54 & 1.07 \\
$10^8$ & 0.62 & 0.24 & 0.67 & 0.68 \\
$10^{10}$ & 0.63 & 0.24 & 0.67 & 0.80 \\
$10^{12}$ & 0.32 & 0.22 & 0.39 & 0.82 \\
\colrule
Average: & 0.51 & 0.23 & 0.57 & 0.84 \\
\botrule
\end{tabular}}
\label{corr}
\end{table}

We have calculated the temporal and spatial correlation 
scales, defined at the point where the normalised autocorrelation function drops from 1 to 0.5 
for the first time, from the same data modelling the 2D Rayleigh-B\'enard convection as used
in the machine learning approaches. We have employed the calculation techniques detailed in 
\cite{saxton24} to calculate these correlation scales. The results of this analysis are shown
in Table \ref{corr}. Perhaps unsurprisingly, all of these scales seem relatively constant with 
respect to $R$ and we include the average values across $R$ in the bottom row of the table; this shows that the correct non-dimensionalisation can yield results that are not sensitive to parameter values.
The PDFs of these values across the domains are relatively narrow, especially for $x_{corr}$.
It should be noted that this does support our finding that the same optimal pySINDy hyperparameter 
set of (12,10,8) worked across varying $R$. Dividing the domain size for the machine learning
analysis ($xdom$,$zdom$,$tdom$) = (4,1,4) by the optimal hyperparameter set (12,10,8), gives the 
physical size of the spatiotemporal subdomains: (0.33,0.1,0.5). This is not in full agreement,
but it can be noted from the range of values explored and limited by memory capacity, we did
not go lower than 8 for $tdiv$. The trend with varying $xdiv$, $zdiv$ and
$tdiv$ in Figure \ref{fig:bulk1} is for the residual error of the machine learning fit 
to improve (decrease) as the values approach these
correlation scales. The comparison with the SPIDER spatiotemporal subdomains is less obvious, as
the physical sizes of the boxes varied in $z$, but we were typically using $\Omega_i$ equal to 
128 or 64 for 2D Rayleigh-B\'enard convection, corresponding to 0.5 or 0.25 in $x$ and 1.28 or
0.64 in $t$, which are pleasingly close to the average correlation time and spatial scales in both cases.
On this basis we would suggest that hyperparameter exploration takes 
the correlation time and spatial scales of the bulk flow as a default 
starting point for the size of the spatiotemporal subdomains. It may also be 
necessary to consider the correlation scales in the boundary layer to infer
the small terms that scale as $R^{-0.5}$. 

\cite{gure19} considered the physical and mathematical intuition regarding
optimal integration domain size. They found that optimal integration domain 
size is a balance. Too small a domain and the error is large because the integration 
domain is too small to effectively average out the influence of noise. Further,
the numerical quadrature error becomes large. Too large a domain and the machine
learning analysis enters a regime where the error should grow exponentially in the dimensions
$H_x$, $H_y$, $H_z$ and 
$H_t$ of the integration domains. 
The optimal choice is a crossover between these two regimes, which we find herein 
has a similarity to the correlation scales. The interested reader may also want to
consider the recent Handbook of Numerical Analysis Vol. 25, and in particular 
Chapter 2 \citep{bortz24}.

\subsection{Numerical accuracy}

\begin{figure}
\begin{center}
\subfigure[$R=10^6$]{
\resizebox*{8cm}{!}%
{\includegraphics[width=0.45\linewidth]{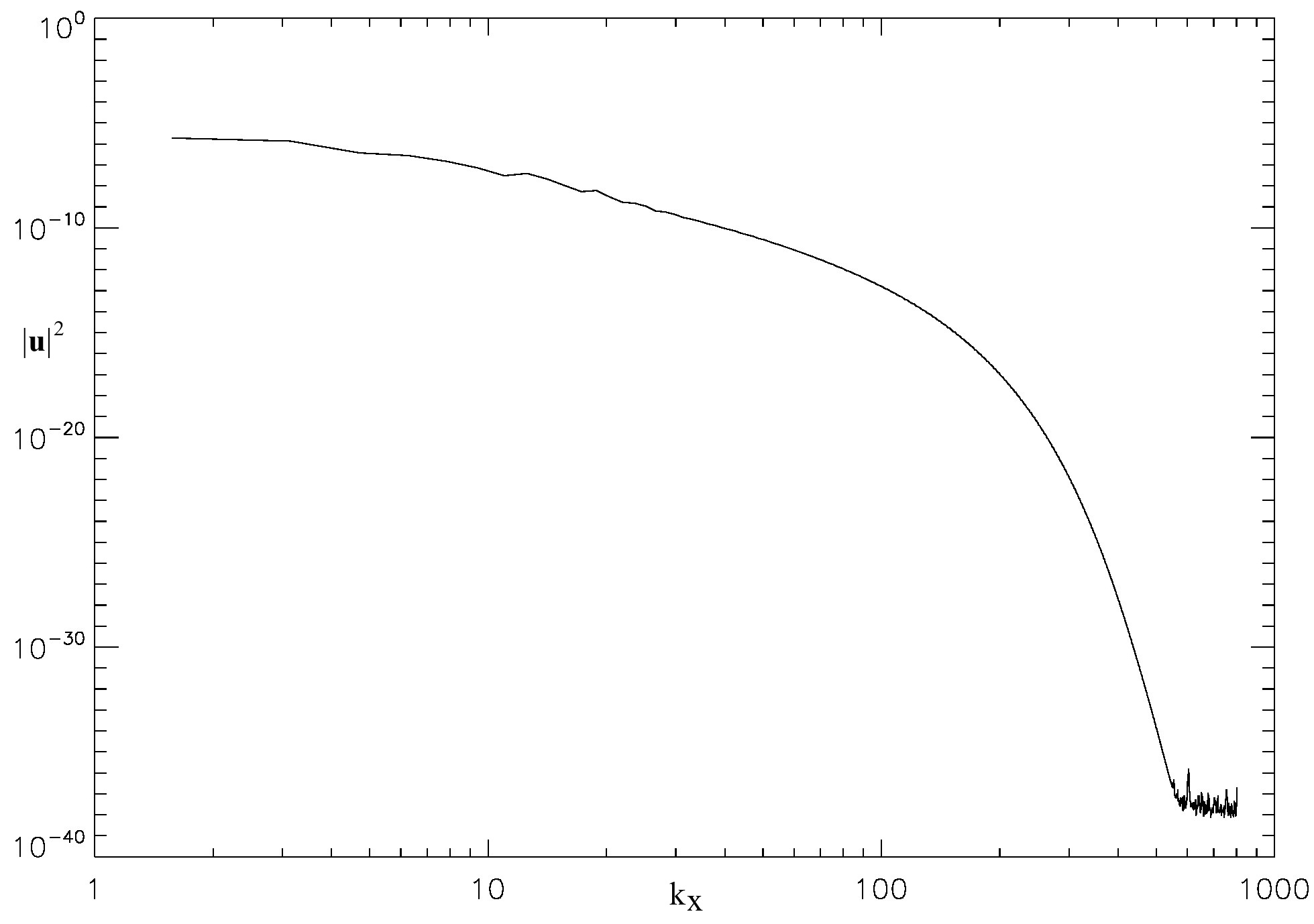}}}%
\subfigure[$R=10^8$]{
\resizebox*{8cm}{!}%
{\includegraphics[width=0.45\linewidth]{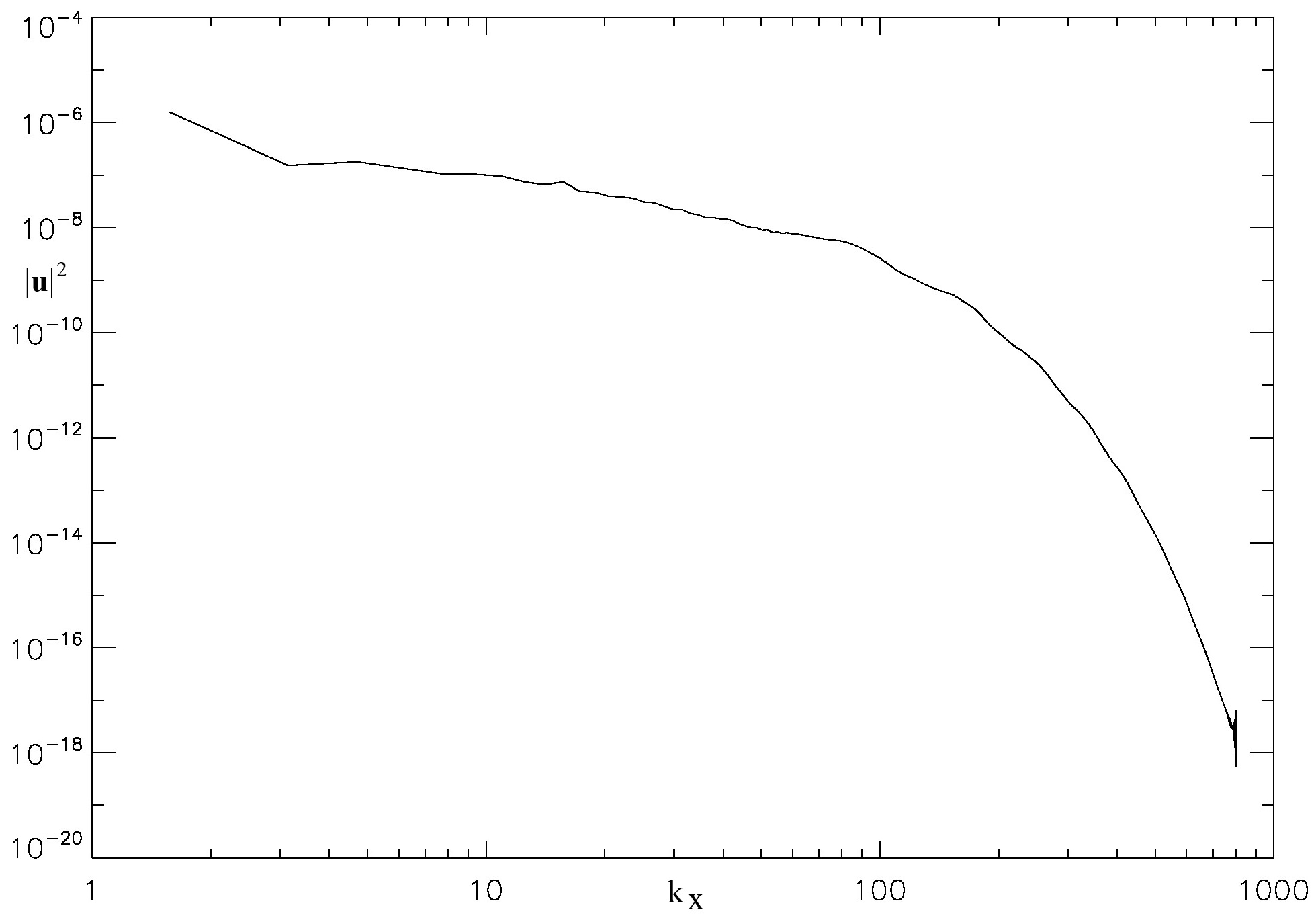}}}\\
\subfigure[$R=10^{10}$]{
\resizebox*{8cm}{!}%
{\includegraphics[width=0.45\linewidth]{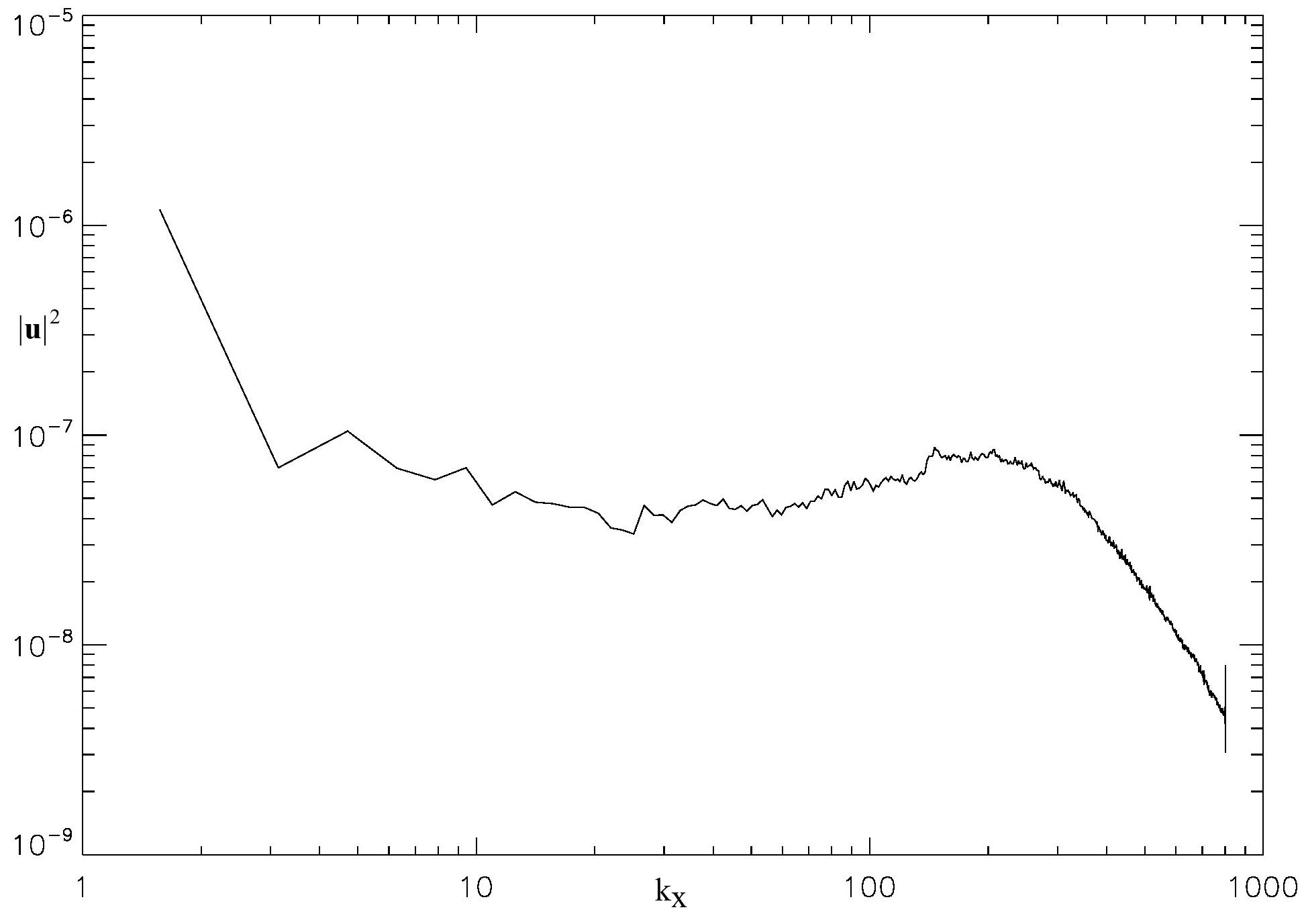}}}%
\subfigure[$R=10^{12}$]{
\resizebox*{8cm}{!}%
{\includegraphics[width=0.45\linewidth]{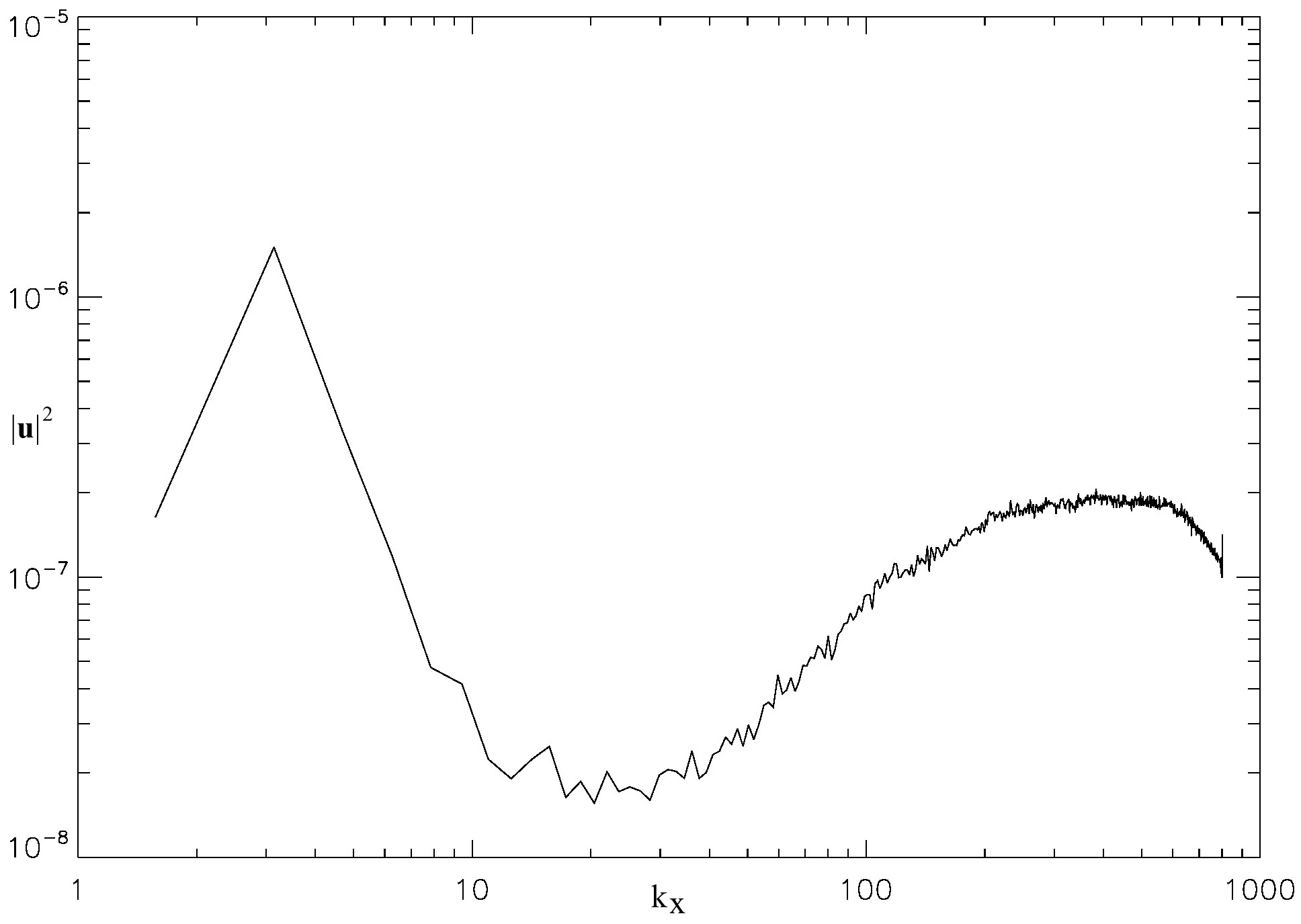}}}%
\caption{
1D power spectra along profiles in the $x$ direction ($k_x$ only)
at a height of 10 Chebyshev nodes above the boundary layer for the
2D simulations of Rayleigh-B\'enard convection, so as to sample
the power spectra of the small-scale dissipative structures in the
boundary layer. As can be seen from
the top row, the complete power spectra of the boundary layer is 
captured at Rayleigh numbers of $10^6$ and $10^8$. As can be seen
from the bottom row, most of the power spectra at $10^{10}$ is
captured, but at $R=10^{12}$ the simulations capture the peak
of power spectra in the boundary layer only.}
\label{figres}
\end{center}
\end{figure}

The failure of the machine learning methods to recover all of the 
terms -- and hence all of the physics -- contained in the original 
governing equations raises the important question of whether this 
is due to a limitation of these methods or the underlying data that 
is used. More precisely, we should question whether that failure is 
indicative of the simulations with the highest values of $R$ being 
sufficiently well resolved. Our choice of Chebyshev polynomials in 
the vertical direction with the number of cells in the boundary 
layer matching published DNS simulations (as discussed in Section 
\ref{nummeth}) and tracking of statistics and monitoring of the 
simulations gave us initial confidence in the quality of the data,
but on the other hand the presence of spurious terms in the equation
recovery is concerning.

In an attempt to definitively answer this question, noting that the 
difficulties have been in the recovery of diffusive terms in 2D
Rayleigh-B\'enard convection, which takes the largest values in the 
boundary layers, we investigated the resolution of the data in that 
region. Since the vertical grid spacing in the boundary layer is 
much finer than the horizontal grid spacing, we focused on the power 
spectra along the $x$ direction at a height just above the boundary.
The results are shown in Figure \ref{figres}. Clearly the boundary layer in the
$R=10^6$ and $R=10^8$ simulations (shown across the top row) is
spatially well resolved, accurately capturing the steep gradient of 
the spectra beyond the peak to many orders of magnitude below peak
power in the boundary layer.
However, at $R=10^{10}$ we see that far less of the
steeply dropping spectrum beyond the peak at $k_x^*\approx 200$ is captured with only 1024
points in the horizontal direction. At $R=10^{12}$, it's now clear
that with the same horizontal resolution, DNS does not capture the  
structures in the boundary layer corresponding to wavenumbers $k_x$ greater than the peak $k_x^*\approx 500$. 
With the smallest scales ($k_x>k_x^*$) unresolved
in the boundary layer, it is not surprising that the both
pySINDy and SPIDER fail or struggle to recover the diffusive terms at
these high $R$. In fact, it is worth emphasizing that SPIDER
was telling us, by recovering Euler rather than Navier-Stokes equations, 
that the DNS data is under-resolved before we confirmed this for the 
boundary layer through these power spectra. Future theoretical
work in particular may therefore be able to use machine-learning tools 
as an additional method of validation and verification of simulations 
to demonstrate that they are indeed accurately resolving everything
the governing equations should be generating in the flow. We strongly 
suggest that any future work explores the resolution question very 
carefully if the machine learning methods are to prove their true 
usefulness in discovering new physics, rather than just demonstrating an
ability, when carefully tuned, to recover known properties of a particular
problem. Spatial resolution can clearly be examined by power spectra not
only of the whole flow and bulk flow, but also of horizontal
slices in the boundary layer. Dedalus also allows the end-user to track
maximum values of the tau terms, which for example relate to the 
divergence of ${\bf u}$, and can be used to establish the necessary
resolution in any direction using a Chebyshev basis, and we have monitored this to ensure
we have resolved the numerical simulations. However, it is not clear that
the choice of the Chebyshev basis, which place most of the nodes near
the boundaries, is necessarily the wisest approach for fully turbulent
simulations where small-scale structure can also be found in the bulk
flow region. The DNS results \citep{zhu18} to which we have compared our 
resolution in Section \ref{nummeth} use the Fourier basis in all directions.
It is also not yet clear, at least from this work, how 
temporal resolution between the `snapshots' that are used in
the machine-learning methods may affect their performance, 
but we suggest this avenue could also be easily explored in future.

\subsection{Guidance for future end-users}

\begin{table}
\tbl{Benefits of either machine learning approach}
{\begin{tabular}{@{}cc}\toprule
pySINDy implementation & SPIDER framework\\
\colrule
Generalisability to further problems & Specific applicability to tensor PDEs \\
Flexibility of library generation$^{\rm a}$ & Specificity of library generation \\
Ease of hyperparameter sweeping using pySINDy & Ease of Matlab scripting \\
Range of optimisers available & Ease of optimiser adjustment \\
Extensive documentation$^{\rm b}$ & Relative high speed \\
Support group$^{\rm c}$ & Interpretable results \\
& Ability to validate and verify DNS\\
\botrule
\end{tabular}}
\tabnote{$^{\rm a}$Although flexibility without guidelines can more of a problem for these complex PDE problems.}
\tabnote{$^{\rm b}$https://pysindy.readthedocs.io/en/latest/}
\tabnote{$^{\rm c}$https://github.com/dynamicslab/pysindy/issues}
\label{proscons}
\end{table}

In an attempt to provide helpful guidance to readers for the future work and the 
application of these methods to other problems, we tabulate some of the advantages of either
method in Table \ref{proscons}.

For our purposes, we have found above that either method is capable of generating results,
although SPIDER has been easier to apply than pySINDy. The memory problems we have observed with
pySINDy may be avoided by a different approach to library generation. The ability to
parameter sweep with SPIDER may be possible with command line application of a reduced 
library. That said, investigation of the results of hyperparameter sweeping requires 
very careful examination of every recovered set of equations in order to check the 
model and find optimum hyperparameter combinations. We are aware that a future Python
version of SPIDER may also make parameter sweeping easier.

The search for the optimal size of spatiotemporal subdomains may be avoided by the detailed
examination of the raw flow data. Optimal parameters for pySINDy clearly show a similarity
to the correlation lengths and times we have calculated for the simulations of 2D
Rayleigh-B\'enard convection. Further investigation may then be able to more thoroughly
explore the variation of the order of the integrating polynomial $p$ for the pySINDy implementation
and the same thing in SPIDER, the power of the weight function $\beta$. We have found that
starting from 8 and exploring the range from 4 to 15 is worthwhile. That said, the 
uniformity of the gridding can strongly affect the performance on variation of this
parameter and conversion of non-uniform grids to uniform grids is worth consideration.
Note though that SPIDER's advantage herein may in part be due to effect of the 
non-uniform vertical grid spacing allowing for resolution of the boundary layers.
Finally, it bears repeating that variation of the MIOSR tolerance in the pySINDy implementation of SINDy 
and in fact any tolerances for these optimisers requires further investigation.

\section{Summary and future work}\label{conc}

In this work, we have applied two machine learning methods to simulated convective
fluid flows. In particular, we have applied the Sparse Identification of Nonlinear 
Dynamics (SINDy) algorithm through the pySINDy implementation and the Sparse Physics-Informed Discovery of Empirical 
Relations (SPIDER) framework with the aim of equation recovery from the raw data.
Both pySINDy and SPIDER have proved able to recover governing equations of the 
2D Rayleigh-B\'enard convection flow simulated.
SPIDER was also able to recover governing equations of the
3D Rayleigh-B\'enard convection flow and planar convective Couette flow simulated.
All simulated data was generated using the Dedalus PDE framework. 
However, our method of 
application has shown that whilst pySINDy and SPIDER are essentially doing the same 
thing, it is clear that the generalised approach of pySINDy puts it at a 
disadvantage when examining problems that SPIDER is tailored to from the outset. 
Specifically, the creation of suitable libraries of possible terms for selection 
in the sparse regression process is physically intuitive with SPIDER as opposed
to combinatorially brute force approach-like with pySINDy. This generation of large
libraries with pySINDy followed by the deactivation of many terms in the library due
to physical constraints is likely to be presenting itself in the high memory demands 
of the algorithm as we have applied it, 
so much so that we were only able to apply pySINDy to the recovery of the governing
equations of 2D Rayleigh-B\'enard convection at Rayleigh numbers of
$10^6$, $10^8$ and $10^{10}$. Sweeping of the hyperparameters used
to perform the sparse regression, specifically the size of spatiotemporal subdomains,
was possible though and analysis indicates that properties of the flow, specifically 
correlation lengthscales and correlation timescale should inform the initial selection
of these parameters. With SPIDER, we were able to demonstrate the recovery of governing
equations, constraints (the incompressibility condition) and boundary conditions for
all the fluid flows considered: 2D Rayleigh-B\'enard convection at Rayleigh
numbers of $10^6$, $10^8$, $10^{10}$ and $10^{12}$ (albeit with performance
limited at $R=10^{12}$ most likely for reasons of numerical accuracy which
it seems SPIDER can robustly identify), 
3D Rayleigh-B\'enard convection at Rayleigh numbers of $10^4$, $10^5$, 
$10^6$ and $10^7$, and 2D plane convective Couette flow at a Rayleigh number
of $10^8$ with varying moving boundary velocities.
In combination with the fact that SPIDER is able to
directly recover scalar {\it and vector} equations, it is clear that it is well-tailored
to the search for potential new turbulence closures with 
the addition of extra library terms (e.g. cubic and quartic derivatives) and
higher rank tensor equations,
something that we intend to examine in future work as development of
these methods allows such capabilities. 
A harder test of both methods may have been to include such higher derivatives
in both libraries and perhaps it is little surprise that a method performs
better when more knowledge is applied in the generation of the library. 
All the same, we would recommend starting
with the SPIDER framework for similar problems to those considered here and using flow
properties, specifically the time and spatial correlation scales, to inform the initial
selection of spatiotemporal subdomain sizes. Investigation of equation recovery can then
focus on the importance of other hyperparameters, such as the order of the integrating
polynomial in pySINDy and the equivalent power of the weighting function $\beta$ in SPIDER, as well
as tolerance of the optimising methods. We make available
all our Dedalus, pySINDy and SPIDER scripts in the hope that they prove useful to anyone
considering this or similar problems in the future.

In conclusion, we look forward to some kind of automation procedure for the pruning
and constraining of candidate term libraries for the recovery of governing PDEs, as
well as further intuition that can be garnered from the raw data to inform machine
learning, as both of these appear key to streamlining our methods and applying them
to more complex problems. We also anticipate that the use of such machine
learning methods may now open a new avenue to validate and verify DNS
in both a qualitative sense (are the equations being solved reconstructed?)
and a quantitative sense (what is the residual of the equations which does
not rely on a particular discretization scheme?).

\section{Acknowledgements}

This project has received funding from
the European Research Council (ERC) under the European
Union’s Horizon 2020 research and innovation programme
(Grant agreement No. D5S-DLV-786780). The calculations
for this paper were performed on the University of Leeds
ARC4 facility, hosted and enabled through the ARC HPC
resources and support team at the University of Leeds, to
whom we extend our grateful thanks. We express our thanks
to Dr Curtis J Saxton for the application of his correlation
length calculation scripts to our data.

\section{Declaration of interests}

The authors are unaware of any conflict of interest.

\section{Data \& Materials Availability}

With the intention of making this entire work reproducible
and applicable by other researchers to different data in the future, all 
of the code, model initialisations, libraries, data,  
instructions for the recreation of the larger
datasets and machine learning analysis scripts are available
in a data repository 
accompanying this paper\footnote{https://doi.org/10.5518/1577}
provided by the University of Leeds Research Data Repository Service.

\section{Author contributions}

All authors contributed to the conception and design of this work.
Data generation and analysis were performed by Chris Wareing. Initial
pySINDy guidance and scripting was co-developed with Alasdair Roy.
Initial SPIDER guidance and scripting was co-developed with Matt Golden
and Roman Grigoriev. The first draft of the manuscript was written by
Chris Wareing and all authors helped with manuscript further development
and revision. Steven Tobias provided critical leadership and guidance.
All authors read and approved the final manuscript.

\end{document}